\documentclass[sn-mathphys-ay,margin=0.5in]{sn-jnl}

\usepackage{graphicx}%
\usepackage{multirow}%
\usepackage{amsmath,amssymb,amsfonts}%
\usepackage{amsthm}%
\usepackage{mathrsfs}%
\usepackage[title]{appendix}%
\usepackage{xcolor}%
\usepackage{textcomp}%
\usepackage{manyfoot}%
\usepackage{booktabs}%
\usepackage{algorithm}%
\usepackage{algorithmicx}%
\usepackage{algpseudocode}%
\usepackage{listings}%
\usepackage{pdflscape}
\usepackage{rotating}

\usepackage{bm}
\usepackage{longtable}
\usepackage{caption}
\usepackage{subcaption}
\usepackage{setspace}
\usepackage{rotating}
\usepackage[percent]{overpic}
\usepackage{cleveref}

\theoremstyle{thmstyleone}%
\theoremstyle{thmstyletwo}%

\theoremstyle{thmstylethree}%

\begin{document}

\title[Article Title]{Parameter identifiability of a neuroendocrine-inflammatory model across experimental designs}


\author[1,2]{\fnm{Aubrey} \sur{Ayres}}\email{amayres@ncsu.edu}

\author[1]{\fnm{Mitchel} \sur{Colebank}}\email{mjcolebank@sc.edu}

\affil*[1]{\orgdiv{Department of Mathematics}, \orgname{University of South Carolina}, \orgaddress{ \city{Columbia}, \state{South Carolina}, \country{US}}}

\affil[2]{\orgdiv{Department of Mathematics}, \orgname{North Carolina State University}, \orgaddress{ \city{Raleigh}, \state{North Carolina}, \country{US}}}


\abstract{Mathematical modeling helps us identify and investigate complex mechanisms of function. This is especially useful for understanding inflammation, which is a complex, multiscale process that interacts nonlinearly with other physiological systems. The full potential of these models can only be realized when combined with experimental data through parameter estimation. However, uniquely determining model parameters requires that they are practically identifiable, which can be assessed by multiple, possibly inconsistent, methods. Identifiability can also vary with experimental designs and observation operators. This study investigates this issue by assessing parameter identifiability in a coupled model of the neuroendocrine-inflammatory system. We compare three workflows following a global sensitivity analysis: Fisher-information matrix based identifiability, global optimization with frequentist confidence intervals, and the profile-likelihood. We compare results across six observation operators with and without measurement noise, all of which are experimentally feasible given prior literature. Our results show that Fisher-information-based methods typically lead to larger sets of identifiable parameters while profile-likelihood analyses lead to a smaller number of identifiable parameters. This suggests that, while computationally expensive, the profile-likelihood is necessary for accurately assessing identifiability, and must be recalculated when the experimental design is altered.}

\keywords{Practical Identifiability, Sensitivity Analysis, Experimental Design, HPA-axis, Immune system}



\maketitle

\section{Introduction}\label{sec:intro}

The immune response to external pathogens is a complex physiological process that interacts across different systems of the body \citep{bangsgaard_integrated_2017,chow_acute_2005}. When the immune system is challenged, the first line of defense is acute inflammation. Inflammatory challenges, such as the use of endotoxin, also known as Lipopolysaccharide (LPS), can be used in humans to study the innate immune response. LPS causes phagocytic cells at the site of injection to recognize endotoxin and respond by inducing inflammation. While inflammation is acutely a protective process, it can also cause damage to tissues long-term. Therefore, the response must be tightly regulated to produce the desired effect of protecting the body without causing unnecessary damage. The coordinated release of cytokines initiates the immune response, hence we are interested in cytokines as pro-inflammatory (induces inflammation) or anti-inflammatory (reduces inflammation). Major cytokines include Tumor Necrosis Factor alpha (TNF-$\alpha$), a pro-inflammatory cytokine, as well as anti-inflammatory cytokines Transforming Growth Factor beta (TGF-$\beta$) and Interleukin-10 (IL-10). 

The Hypothalamic-Pituitary-Adrenal (HPA)-axis is a neuroendocrine pathway involved in the human stress response, ultimately regulating the release of stress hormones and glucocorticoids. The main HPA actors are corticotropin-releasing hormone (CRH), adrenocorticotropic hormone (ACTH), and cortisol, the latter of which is the main stress hormone released as a response to both physical and mental stress. It is also consistently produced by and present in the body to maintain basic homeostasis. The HPA and immune systems interact to regulate physiological function. TNF-$\alpha$, e.g., connects the immune response to the HPA-axis by stimulating CRH release, followed by increased ACTH and then cortisol release. Cortisol inhibits both CRH and ACTH, forming a closed feedback loop that’s meant to keep all three hormones in balance \citep{bangsgaard_integrated_2017}. HPA dynamics are coupled to the circadian rhythm, and also fluctuate on smaller time-scales following ultradian rhythms. The interplay between the immune response and hormonal cycling in the body presents complex oscillatory dynamics that can change with even slight perturbations \citep{chow_acute_2005}. For instance, long-term stress in the form of post traumatic stress disorder (PTSD) leads to chronic overstimulation of the HPA-axis and changes to cortisol sensitivity, leading to dysregulation of inflammation \citep{somvanshi_role_2020}.


Mathematical models and the integration of experimental data provide a synergistic tool for describing these complex processes. A variety of models exist for describing the acute inflammatory response and the HPA-axis as separate systems. Reynold's et al. developed a 4-state model that is governed by anti-inflammatory mediators \citep{reynolds_reduced_2006}. Baker et al. considered the concentrations of both pro- and anti-inflammatory cytokines in their model of the inflammatory response in rheumatoid arthritis patients \citep{mbaker}. Chow et al. developed a comparatively complex model to examine inflammation in varying shock states \citep{chow_acute_2005}. Modeling the HPA-axis system requires capturing precise circadian and ultradian rhythms, which necessitate more complex model equations. Vinther et. al use a minimal 3-state model to simulate ultradian rhythms in the HPA-axis \citep{vinther_minimal_2011} while Meyer-Hermann et al. developed a reduced view of the immune system in their model for the HPA-axis, including the effect of TNF-$\alpha$ on cortisol levels \citep{meyerhermann}. Malek et al. consider the bidirectionality of the two-systems to be a focal point, employing delayed equations that take the immune system and HPA-axis into account to produce accurate dynamics \citep{malek}. Relevant to our work, Bangsgaard et al. expand on this model further by using nonlinear ODEs to describe the coupled systems, significantly reducing the dimensionality present in the model by Malek et al. \citep{bangsgaard_integrated_2017}. The result is coined the Integrated Inflammatory Stress (ITIS) model, which simulates circadian and ultradian rhythms, as well as dynamics over the course of multiple days. 



To fully realize a model's potential, the model should be calibrated to data by solving an inverse problem \citep{cintron-arias_sensitivity_2009}. Parameters that are identifiable given the data will produce unique model outputs for unique inputs (analogous to the notion of an injective function). Identifiability is classifed into two main categories. Structural identifiability is dependent on the mathematical structure of a model, and is a prerequisite for practical identifiability \citep{simpson_parameter_2026}. Practical identifiability is concerned with experimental design and measurement error. A system may be structurally identifiable, but insufficient availability or quality of data can cause parameters to become practically non-identifiable. Parameters can be practically non-identifiable for a variety of reasons, including (i) being noninfluential on the model output, or (ii) being involved in interactions with other parameters that lead to confounding estimates. In the former case, we eliminate the parameter (via setting it to a constant, nominal value) to reduce unnecessary complexity, whereas the latter case requires similar fixing or changing the model's structure, which can be analytically intensive \citep{smith_uncertainty_2014}. 

Practical identifiability can be further attributed to data, which can be noisy or not sufficiently available. Noise-free data may present identifiability issues if there is not a sufficient number of data points or if data are not observed in an informative manner (e.g., when dynamics are not sufficiently captured). In either case, the data fails to provide enough information to characterize the output space and infer parameters accurately. These effects are most often lumped together, yet may be distinct and separately ask whether the experimental design is informative versus whether the technology for measuring data is sufficiently accurate. Derivative based sensitivity methods provide an easy first step in investigating identifiability \citep{taylor-lapole_parameter_nodate,colunga_parameter_2023,cintron-arias_sensitivity_2009}, but approximate the underlying identifiability up to first-order accuracy at a local point. These methods extend into structural correlation and orthogonal sensitivities methods for testing identifiability \citep{olsen_parameter_2019}. Global sensitivity methods capture nonlinear behavior and interactions between parameters\citep{colebank_assessing_2025,taylor-lapole_parameter_nodate,colunga_parameter_2023} and can better assess how influential parameters are. However, there is no direct link to identifiability aside from identifying parameters that have no influence on the quantity of interest \citep{smith_uncertainty_2014}.  The profile-likelihood has emerged as a gold standard for overcoming issues with sensitivity-based methods \citep{simpson_parameter_2026,colebank_assessing_2025}, as well as the use of multi-start inference to determine how unique parameter estimates are \citep{colunga_parameter_2023}. 

This manuscript investigates some of the independent effects of the causes of practical identifiability across a range of analytical methods. In doing so, we may compare the conclusions and interpretations of these analyses in a variety of conditions. Our analysis centers around the ITIS model as a prime example of a biological model that requires rigorous identifiability analysis, due to the possible variations in experimental design that exist in practical application. We select four commonly applied analysis methods varying in their difficulty to implement and cost to compute. 

\section{Mathematical Model}
\begin{figure}[h!]
    \centering
    \includegraphics[width=0.8\linewidth]{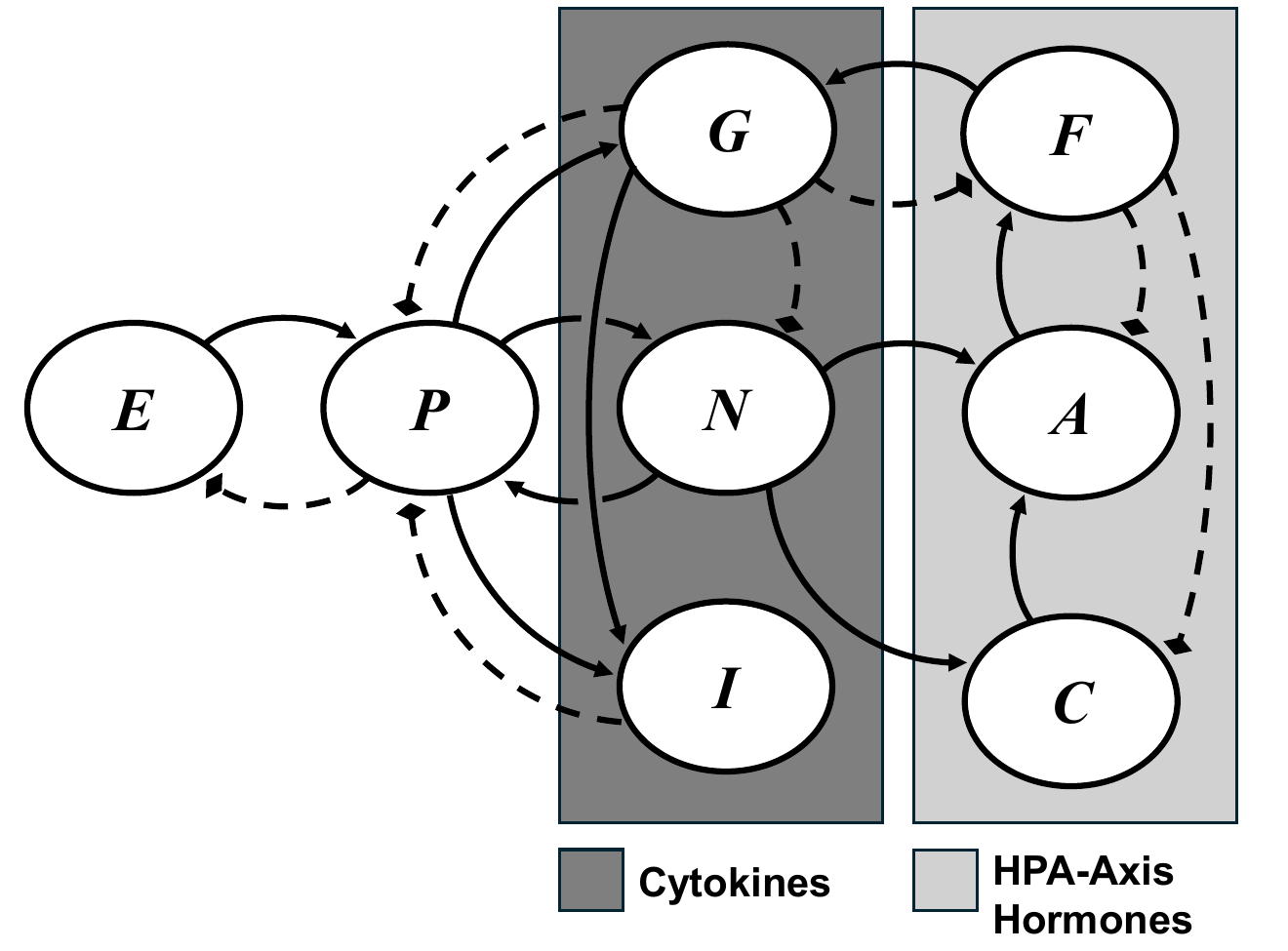}
    \caption{Schematic describing the relationships between the immune system cytokines and HPA-axis hormones. Solid arrows represent stimulating interactions, and dashed arrows represent inhibitory interactions}
    \label{fig:ITIS_Schematic}
\end{figure}

The ITIS model is a system of nonlinear ordinary differential equations that couples the inflammatory response to the HPA-axis. The model system is depicted in \cref{fig:ITIS_Schematic}. It has eight states: Endotoxin (\textit{E}), Activated Phagocytic Cells (\textit{P}), TGF-$\beta$ (\textit{G}), TNF-$\alpha$ (\textit{N}), IL-10 (\textit{I}), CRH (\textit{C}), ACTH (\textit{A}), and Cortisol (\textit{F}). $R(t)$, an external function that simulates the influence of the circadian rhythm of a 24 hour cycle. There are 43 parameters: Parameters $d_i$ represent the rate at which the body removes the state from the system, $k_j$ represent the strength of stimulation or inhibition, $h_l$ are half-saturation constants, $b_m$ are basal production rates, and the remaining parameters govern the circadian rhythm function $R(t)$.

We summarize the relationships between the model states as follows:
\begin{itemize}
    \item \textbf{Endotoxin (\textit{E})} - Introduced to the model via initial conditions, eliminated by phagocytic cells (\textit{P}). \\
    \item \textbf{Phagocytic Cells (\textit{P})} - Activated by the presence of endotoxin (\textit{E}). This state is up-regulated by pro-inflammatory cytokines (\textit{N}) and down-regulated by anti-inflammatory cytokines (\textit{G} and \textit{I}). \\
    
    \item \textbf{TGF-$\beta$ (\textit{G})} - Released by phagocytic cells (\textit{P}) and up-regulated by CRH (\textit{C}). \\
    
    \item \textbf{TNF-$\alpha$ (\textit{N})} - Released by phagocytic cells (\textit{P}) and inhibited by TGF-$\beta$ (\textit{G}). It also up-regulates its own expression. \\
    
    \item \textbf{IL-10 (\textit{I})} - Produced by phagocytic cells (\textit{P}) and up-regulated by TGF-$\beta$ (\textit{G}). The body also produces a basic level of IL-10 (\textit{I}) when no immune challenge is present. \\
    
    \item \textbf{CRH (\textit{C})} - Release follows a circadian rhythm governed by $R(t)$. A basal level is present in the body. Release is also stimulated by TNF-$\alpha$ (\textit{N}) and inhibited by cortisol (\textit{F}). \\
    
    \item \textbf{ACTH (\textit{A})} - Release is stimulated by CRH (\textit{C}), and this release is inhibited by cortisol (\textit{F}). ACTH release is also stimulated by TNF-$\alpha$ (\textit{N}). \\
    
    \item \textbf{Cortisol (\textit{F})} - Release is stimulated by ACTH (\textit{A}). This release is inhibited by TGF-$ \beta$ (\textit{G}).
\end{itemize}

These relationships are described mathematically by the ITIS model: 
\begin{align}
\frac{dE}{dt} & = -d_1EP \label{eq:ITIS_E}\\
\frac{dP}{dt} & = k_1\left(\left(1+k_2\frac{N}{h_1+ N}\right)\cdot\frac{h_2}{h_2 + G}\cdot\frac{h_3}{h_3 + I}\right)E -d_2P \label{eq:ITIS_P}\\
\frac{dG}{dt} & = k_3P + k_4\frac{C}{h_4 + C} - d_3G \label{eq:ITIS_G}\\
\frac{dN}{dt} & = \frac{P}{h_5 + P} \cdot \frac{h_6^4}{h_6^4 + G^4}\left(k_5 + k_6\frac{N}{h_7 + B}\right) - d_4N^2 \label{eq:ITIS_N}\\
\frac{dI}{dt} & = b_1 + k_7\frac{P^3}{h_8^3+P^3} + k_8\frac{G^6}{h_9^6 + G^6} -d_5I\frac{h_{10}}{h_{10} + I} \label{eq:ITIS_I}\\
\frac{dC}{dt} & = b_2 + k_9R(t)\frac{C}{1+k_{10}F^2} + k_{11}N - d_6C \label{eq:ITIS_C}\\
\frac{dA}{dt} & = k_{12}\frac{C}{1+k_{13}F} + k_{14}\frac{N^2}{h_{11}^2 + N^2} - d_7A \label{eq:ITIS_A}\\
\frac{dF}{dt} & = k_{15}\frac{A^2}{1+k_{16}G} - d_8F \label{eq:ITIS_F}\\
R(t) &= N_c\left(\frac{t^k_m}{t^k_m + \alpha^k} \cdot \frac{(T_t - t_m)^l}{(T_t - t_m)^l + \beta^l}\right) \label{eq:ITIS_R}
\end{align}

To run our simulations, we set $E=0$ to simulate homeostasis with no immune challenge or stressor for a 72 hour-period in the model. Then, we set $E=2.0$ (ng/kg) to simulate an injection of endotoxin, which triggers an immune response similar to that used by Bangsgaard et. al (\citeyear{bangsgaard_integrated_2017}). Our focus is solely on the post-injection dynamics and the return to steady state, hence each simulation begins with a 72 hour homeostatic period after which we introduce $E$ to the system and record the output. 
Nominal parameter values as provided by Bangsgaard et al. can be found in Table \ref{table:1} (\citeyear{bangsgaard_integrated_2017}).

\begin{table}[ht!]
    \centering
    \renewcommand{\arraystretch}{1.3}
\begin{tabular}{c c c || c c c}
Parameter & Nominal Value & Unit & Parameter & Nominal Value & Unit\\
\hline
$d_1$ & $1.35 \times 10^{-7}$ & $(\text{h N-unit})^{-1}$ 
& $h_{10}$ & $791.27$ & $\frac{\text{pg}}{\text{mL}}$ \\

$k_1$ & $4.9956 \times 10^{7}$ & $\frac{\text{N-unit}\cdot\text{kg}}{\text{h}\cdot\text{pg}}$ 
& $b_2$ & $0.001$ & $\frac{\text{pg}}{\text{mL}\cdot\text{min}}$ \\

$k_2$ & $12.94907$ & -- 
& $k_9$ & $6.8400 \times 10^{9}$ & $\frac{\text{pg}}{\text{mL}\cdot\text{min}}$ \\

$h_1$ & $1693.9509$ & $\frac{\text{pg}}{\text{mL}}$ 
& $k_{10}$ & $1.7558 \times 10^{9}$ & $\left(\frac{\text{dL}}{\mu\text{g}}\right)^2$ \\

$h_2$ & $0.07212$ & $\frac{\text{pg}}{\text{mL}}$ 
& $k_{11}$ & $0.0667$ & $\text{min}^{-1}$ \\

$h_3$ & $147.68$ & $\frac{\text{pg}}{\text{mL}}$ 
& $d_6$ & $0.032$ & $\text{min}^{-1}$ \\

$d_2$ & $0.1439$ & $\text{h}^{-1}$ 
& $k_{12}$ & $2.3688 \times 10^{4}$ & $\text{min}^{-1}$ \\

$k_3$ & $0.1546 \times 10^{-8}$ & $\frac{\text{mL}}{\text{pg}\cdot\text{N-unit}\cdot\text{h}}$ 
& $k_{13}$ & $1.7778 \times 10^{5}$ & $\frac{\text{dL}}{\mu\text{g}}$ \\

$k_4$ & $0.5$ & $\frac{\text{mL}}{\text{pg}\cdot\text{h}}$ 
& $k_{14}$ & $112$ & $\frac{\text{pg}}{\text{mL}\cdot\text{min}}$ \\

$h_4$ & $500$ & $\frac{\mu\text{g}}{\text{dL}}$ 
& $h_{11}$ & $80$ & $\frac{\text{pg}}{\text{mL}}$ \\

$d_3$ & $0.031777$ & $\text{h}^{-1}$ 
& $d_7$ & $0.016$ & $\text{min}^{-1}$ \\
\hline
$h_5$ & $550 \times 10^{4}$ & $\text{N-unit}$ 
& $k_{15}$ & $5.0746 \times 10^{-4}$ & $\frac{\mu\text{g}/\text{dL}}{\text{min}(\text{pg}/\text{mL})^2}$ \\

$h_6$ & $0.1589$ & $\frac{\text{pg}}{\text{mL}}$ 
& $k_{16}$ & $12$ & $\frac{\text{pg}}{\text{mL}}$ \\

$k_5$ & $25.5194$ & $\frac{\text{pg}}{\text{mL}\cdot\text{h}}$ 
& $d_8$ & $0.0266$ & $\text{min}^{-1}$ \\

$k_6$ & $3.5514 \times 10^{4}$ & $\frac{\text{pg}}{\text{mL}\cdot\text{h}}$ 
& $\alpha$ & $300$ & $\text{min}$ \\

$h_7$ & $1.5495 \times 10^{3}$ & $\frac{\text{pg}}{\text{mL}}$ 
& $k$ & $5$ & -- \\

$d_4$ & $0.0307$ & $\frac{\text{mL}}{\text{pg}\cdot\text{h}}$ 
& $\beta$ & $950$ & $\text{min}$ \\

$b_1$ & $1187.2$ & $\frac{\text{pg}}{\text{mL}\cdot\text{h}}$ 
& $l$ & $6$ & -- \\

$k_7$ & $267,480$ & $\frac{\text{pg}}{\text{mL}\cdot\text{h}}$ 
& $\epsilon$ & $0.01$ & -- \\

$h_8$ & $8.0506 \times 10^{7}$ & $\text{N-unit}$ 
& $\delta$ & $76.37$ & $\text{min}$ \\

$k_8$ & $43875$ & $\frac{\text{pg}}{\text{mL}\cdot\text{h}}$ 
& $T$ & 1440 & min \\

 $h_9$ & .38 & $\frac{\text{pg}}{\text{mL}\cdot\text{h}}$ & $N_c$ & 1.9168 & -- \\

$d_5$ & 98.932 & $h^{-1}$ & & &\\

\end{tabular}
\caption{Parameter values and units assembled from estimates, data, and previous studies by Bangsgaard et al. (\citeyear{bangsgaard_integrated_2017}). These provide the nominal parameters, $\bm{\theta}_0$, for data generation.}
\label{table:1}
\end{table}

\section{Data and Experimental Design}\label{sec:Data}
Models focused on the HPA-axis have used ACTH and cortisol measurements to inform their model. It is infeasible to measure CRH, especially on a time series \citep{spencer_users_2017}. Current methods allow cortisol and ACTH concentrations to be measured to a relative degree of accuracy every half-hour for a period of hours \citep{spencer_users_2017}. Beyond this, practical limitations exist for how often measurements may be taken. For this reason, dynamics may be preemptively derived from animal models \citep{bangsgaard_integrated_2017}. Limitations exist in this avenue as well, as neuroendocrine data taken from animals can be easily confounded by the stress response induced by data collection \citep{spencer_users_2017}. Many cytokines can be measured using ELISA panels \citep{chiswick_detection_2012}. All of these measurements will come with some level of noise, which must be taken into consideration when calibrating models. 

In models that couple the immune system and HPA-axis, ACTH and cortisol levels are often used in tandem with certain cytokine measures \citep{mbaker,vinther_minimal_2011,chow_acute_2005,bangsgaard_integrated_2017}. TNF-$\alpha$ and IL-10 are commonly obtained among many others. We will choose to consider these four states in our model designs, seeing as they are most commonly considered among the selected papers. 
 
We use these observations of data availability in similar work as well as technical limits to inform our choice of model designs. We therefore find it both reasonable and valuable to investigate the following designs:
\begin{align*}
D^m_1 &: F + A,  &m=13,25\\
D^m_2 &: F + A + N,  &m=13,25\\
D^m_3 &: F + A + N + I,  &m=13,25
\end{align*}
where the superscript $m$ denotes the number of observed time-points in each design. This corresponds to data taken (i) every other hour and (ii) every hour over a period of 24 hours \citep{bangsgaard_integrated_2017}, providing 13 and 25 data points, respectively. We examine noise-free and noisy data as discussed later. In the latter, we assume that $\sigma_\epsilon = 4$ across all state variables when data  is used. 


\section{Methods}
Consider the general problem
\begin{equation}
    y_{ij} = \mathcal{H}\left(\bm{u}(t_i;\bm{\theta})\right)+\epsilon_{ij}
\end{equation}
where $y_{ij}\in\mathbb{R}$ is the observation for state $j$ at time point $t_i$. These observations are then linked to the state variables $\bm{u}:\mathcal{T}\times\Theta\mapsto\mathcal{U}$ on the time domain $\mathcal{T}$ and parameter space $\Theta$, where the observation operator $\mathcal{H}:\mathcal{U}\mapsto\mathbb{R}$ maps the specific state variable and time point to the observations available. We assume the measurement errors are corrupted by independent and identically distributed (iid) Gaussian noise, $\epsilon_{ij}\sim\mathcal{N}(0,\sigma_{\epsilon}^2)$, with measurement error variance $\sigma_{\epsilon}^2$. Our model is governed by systems of nonlinear ODEs, where $\bm{u}$ corresponds to the inflammatory and neuroendocrine states defined in eqs.~\eqref{eq:ITIS_E}-\eqref{eq:ITIS_F}. Our general mathematical question is whether the parameters exhibit \textbf{practical identifiability}. That is, can we uniquely determine a subset of parameters $\bm{\theta^*}\in\bm{\Theta^*}\subset\bm{\Theta}$ given the observation operator $\mathcal{H}$ and the measurement error $\epsilon_{ij}$. Given that our parameters in Table \ref{table:1} vary in magnitude but are strictly positive, we consider the log-transformed parameters $\bm{\tilde{\theta}}=\ln(\bm{\theta})$ and strictly-positive domain $\bm{\tilde{\Theta}}=\ln\left(\bm{\Theta}\right)$.

We begin by applying a global sensitivity analysis (Morris Screening) to screen for noninfluential parameters. From this, we obtain a subset of parameters that are deemed non-influential on the observed states. We then compare three different methods for assessing identifiability. The first uses an approximation of the Fisher Information Matrix (FIM) to determine whether parameters are locally identifiable. The second uses a global-optimization approach to determine how parameters converge to a single solution. We also compute the FIM in this second approach, but do so to construct asymptotic confidence intervals based on the central limit theorem \citep{cintron-arias_sensitivity_2009}. Finally, we compute profile-likelihood derived confidence intervals \citep{colebank_assessing_2025,simpson_parameter_2026} as a gold-standard comparison to the prior two methods. We apply these three methods across designs $D^m_1$, $D^m_2$, and $D^m_3$ by modifying the observation operator $\mathcal{H}$ in each setting. 
\subsection{Global Sensitivity Analysis} \label{sec:GSA}
Global sensitivity analysis determines the influence of parameters on the quantities of interest by sampling $\bm{\theta}$ from its parameter space \citep{colunga_parameter_2023,smith_uncertainty_2014}. This is in contrast to local analyses which assess model sensitivity around a nominal value of $\bm{\theta}$. We specifically apply Morris screening to filter out noninfluential parameters that in theory are less informative for inference. 



Morris screening involves the computation of ``elementary effects", the relative change of the model output with respect to a change in parameter value \citep{smith_uncertainty_2014}. The effects of $\theta_i$ on the output quantity are classified as  a) negligible overall, b) linear and additive, or c) having nonlinear effects or higher order interactions with other parameters. Prior to this analysis, parameters are mapped from their logarithmic parameter space $\bm{\tilde{\Theta}}$ to the unit hypercube $[0,1]^{N_p}$, where $N_p$ denotes number of parameters analyzed. The elementary effect of the $j$th sample of the $i$th parameter, $\theta^j_i$, on the time-dependent quantity of interest $f(t;\bm{\theta})$ is then computed as
\begin{equation} \label{eq:EE1}
d^j_i(t;\bm{\theta^j}) = \frac{f(t;\bm{\theta^j} + \bm{e}_i \Delta) - f(t;\bm{\theta^j)}}{\Delta}.
\end{equation}
The step size $\Delta$ is chosen from the set $\Delta \in \{1/(\mathcal{M}-1), 2/(\mathcal{M}-1) \dots, (\mathcal{M}-2)/(\mathcal{M}-1) \}$, where $\mathcal{M}$ denotes the number of possible perturbation sizes. To preserve symmetry of the parameter distributions, $\mathcal{M}$ should be even~\citep{smith_uncertainty_2014}. The elementary effects are computed by sampling the full parameter set $K$ times by drawing each parameter from a uniform distribution and then mapping the parameter back to its original value before evaluating the model. We calculate the 2-norm of the time-dependent elementary effects, $\tilde{d}^j_i(\bm{\theta}) = ||d^j_i(t;\bm{\theta})||_2$ for a scalar metric. Note that the calculation of $d^j_i$ is independent of any observational model or available data.

The elementary effects' mean and variance are obtained by integrating the outcomes from multiple trajectories. The modified Morris' indices are calculated as
\begin{equation} 
\mu^*_i = \frac{1}{K} \sum_{j=1}^K |\tilde{d}_i^j|, \hspace{7mm} \mu_i = \frac{1}{K} \sum_{j=1}^K \tilde{d}_i^j,\hspace{7mm} \sigma^2_i = \frac{1}{K-1} \sum^{K}_{j=1} \left( \tilde{d}_i^j - \mu_i \right)^2,
\label{eq:morris}
\end{equation}
where $\mu^*$ quantifies the magnitude of the individual effect of the input on the output, i.e., the sensitivity of the model with respect to the parameters, while the variance estimate $\sigma^2$ describes the variability in the model sensitivity due to parameter interactions or nonlinearity. The variable $\mu$ is only used to calculate the variance estimate. Parameters with a large $\mu^*$ and $\sigma^2$  have nonlinear or drastic effects on the model output. Here, we consider ranking parameters through the combined metric, $\sqrt{{\mu^*}^2 + \sigma^2}$, as proposed in \citeyear{Wentworth2016}. The randomized Morris algorithm is an efficient algorithm for computing elementary effects and scales the step size $\Delta$ by the parameter magnitude (see Algorithm 2 in \cite{olsen_parameter_2019}), which we use here. 

We assume the parameter space is $\pm20\%$ around the nominal log-scaled value, $\bm{\theta^*}$, for each parameter. The parameter bounds are hence non-symmetric around the true nominal values after exponentiation.  Morris screening is applied to the post-injection models for all eight states. We combine results for these states depending on the experimental design under consideration. To derive a concise metric for the multiple output states in the design, we employ a combined metric to rank parameters based on which model states, $u$, are being considered. The metric is normalized based on the maximum ranking metric for each state in the design
\begin{equation}
    \mathcal{R}^u_i = \frac{\sqrt{{\mu_{u,i}^*}^2 + \sigma_{u,i}^2}}{\underset{i\epsilon\{1:N_p\}}{\max}\left(\sqrt{{\mu_{u,i}^*}^2 + \sigma_{u,i}^2}\right)}.
    \label{eq:rank}
\end{equation}
Setting a threshold for $\mathcal{R}$ is often problem dependent, given that different outputs or combinations of outputs can signficantly impact the quantitative values used for parameter ranking \citep{bangsgaard_integrated_2017,colunga_parameter_2023,dadashova_local_2024}. We set $\mathcal{R}=1/4$ as out parameter cutoff for ``non-influential'' parameters, as this would correspond to, at a minimum, the lower 25\% of parameters. Note that $\max{\mathcal{R}}=|D_i|$, i.e., the sensitivity index can never be larger than the number of state variables, since each sensitivity is normalized to the maximum for a specific state variable.

\subsection{Parameter Estimation}\label{sec:WLS}
We employ a weighted least squares (WLS) approach to calibrate the model to data provided from the experimental designs. Synethetic data are generated by running the model with the nominal parameter values, denoted $\bm{\theta}_0$, provided in Table \ref{table:1}. Then, the state variables corresponding to the designs $D_i^j$ are saved at the specified time-points discussed earlier. We add Gaussian, zero-mean iid noise with $\sigma_\epsilon=4$ as our standard deviation. Given some subset of parameters $\bm{\tilde{\theta}}$ for inference, our goal is to find the minimizer of the weighted sum of squared error, defined as
\begin{equation}\label{eq:WLS}
    \bm{\hat{\tilde{\theta}}} = \underset{\bm{\tilde{\theta}} \in \bm{\tilde\Theta}}{\text{argmin}}\sum_{j=1}^{s}\sum_{i=1}^{n_j}\left(e_{ij}(\bm{\tilde{\theta}})\right)^2, \ \ e_{ij} = \left[\frac{y_{ij} - f_j(t_i;\bm{\tilde{\theta}})}{\bar{y_j}}\right]    
\end{equation}
where $\bar{y_j}$ is the time-series average of the data from state $j$, and $s$ denotes the number of state variables used in the inference procedure. The variable $n_j$ represents the number of sample points available for state $j$. We call $e_{ij}$ the weighted residual. We note that, because the objective function is weighted by the mean of the data, the outputs are nondimensional and closer in magnitude. This is a form of weighted least squares \citep{smith_uncertainty_2014}. We employ the Trust Region Reflective algorithm to minimize cost using the scipy.optimize library.

\subsection{Local Sensitivity Analysis}
The local sensitivity of the WLS residual, $e_{ij}$, of state $j$ and time $t_i$ with respect to log-scaled parameter $\tilde\theta_k$ is
\begin{equation}
    S_{j,k}(t_i)=\frac{d}{d\tilde\theta_k}e_{ij}(\bm{\tilde\theta})  \approx \left(\frac{e_{ij}(\bm{\tilde\theta}+\Delta_k)  - e_{ij}(\bm{\tilde\theta})}{\Delta_k}\right)\theta_k, \ \ \ k\in\{1,2,\dots,N_p\}.
\end{equation}
The latter expression is our finite difference approximation using $\Delta_k$ as the step size. Since the parameters are on a log-scale, we naturally get scaling based on the magnitude of the parameter \citep{olsen_parameter_2019,colunga_parameter_2023}. Carrying out this computation for each parameter $\theta_k$ will result in a sensitivity matrix $\bm{S} \in \mathbb{R}^{N_p \times (N_s \cdot N_t)}$, where $N_s$ is the number of states, $N_t$ is the number of time points, and each row reflects the concatenated vector of residuals across time and state. 

The sensitivity matrix is then used to approximate the Fisher Information matrix $\bm{F}=\bm{S^\top S}$. This matrix contains information about the local concavity of the quantity of interest, and can provide a theoretical lower bound on the covariance matrix of the estimator of the parameters \citep{cintron-arias_sensitivity_2009}. From a linear algebra perspective, $\bm{F}$ provides some limited information about identifiability in terms of linear dependence. The condition number of the matrix, $\textrm{Cond}\left(\bm{F}\right)$, can identify whether the eigenvalues of $\bm{F}$ are vastly different and thus information about whether the parameters are hard to discern \citep{smith_uncertainty_2014,simpson_parameter_2026}. Here, we confirm the identifiability of this selection by following an algorithm developed by Dadashova et. al. (\citeyear{dadashova_local_2024}). Starting with the parameter ranked as most influential, parameters are added back in one at a time. The condition number of $\bm{F}$ is calculated, and if the addition of any parameter causes the condition number to exceed a threshold, the parameter is deemed non-identifiable and removed from the set. We utilize a threshold of $10^6$ in this work as this is a reasonable maximum for determining whether $\bm{F}$ is invertible \citep{colebank2022silico,olsen_parameter_2019}. The local sensitivity analysis is applied to parameters selected in the global analysis, ordered by metric $\mathcal{R}$.

\subsection{Global Optimization with Confidence Intervals}
We consider a second, more global method of assessing identifiability by investigating whether there is a unique, global minimizer for the dataset and parameter subset \citep{colunga_parameter_2023}. For each dataset and parameter subset, we select 20 initializations to then run an optimization routine for each initial guess. The parameters are perturbed according to a relative Gaussian, i.e. $\tilde\theta_i^k\sim\mathcal{N}(\tilde\theta_i,\sigma^2_i)$, where $\tilde\theta_i$ is the data-generating parameter and $\sigma = 0.1\tilde\theta_i$. This perturbs parameters away from their nominal value to then assess whether there are multiple, potentially non-unique local-minima in the cost function. For each initialization, the optimization routine described in Sec. \ref{sec:WLS} is run. 

After the optimization routine, we calculate asymptotic, frequentist confidence intervals for the parameters \citep{cintron-arias_sensitivity_2009,smith_uncertainty_2014}. Under the assumption of Gaussian, independent, additive errors, we can derive the confidence intervals for the parameters in log-space as
\begin{equation}\label{eq:CI_freq}
    \textrm{CI}^{1-a}(\hat{\tilde{\theta_i}}) = \left[ \hat{\tilde{\theta_i}}-t^{1-a}_{N_{_D}-N_p}\textrm{SE}_i,\hat{\tilde{\theta_i}}+t^{1-a}_{N_{_D}-N_p}\textrm{SE}_i \right], \ \ \ \textrm{SE}_i =\sqrt{\sigma_\varepsilon^2 \left(\bm{S^\top S}\right)^{-1}_{ii}}
\end{equation}
where SE represents the standard error derived from the diagonal of the covariance matrix given by $\sigma_\varepsilon^2\left(\bm{S^\top S}\right)^{-1}$ and
\begin{equation}\label{eq:variance_estimate}
    \sigma_\varepsilon^2=\frac{1}{N_{_D}-N_p}\sum_{j=1}^{s}\sum_{i=1}^{n_j}\left(e_{ij}(\hat{\tilde{\theta}})\right)^2, \ \ \ N_{_D} = \displaystyle \left(\sum_{j=1}^s n_j\right) 
\end{equation}
which is calculated from the optimal parameters found, $\hat{\tilde{\theta}}$. The scalar term $N_{_D}$  represents the degrees of freedom for the problem given $s$ output states and $N_p$ parameters. The term $t^{1-a}_{N_D-p}$ is the score from a $t$-distribution at a significance level of $1-a$ and $N_D-N_p$ degrees of freedom.

Then our approach is as follows: given multiple initial parameter guesses, we deduce identifiability issues if parameter confidence intervals derived from eq. \eqref{eq:CI_freq} do not consistently overlap with the true parameter value while also obtaining a nearly identical objective function value. If parameters have relatively large confidence intervals, this may also indicate identifiability issues since a large variance suggests large uncertainty. Tight, bounded confidence intervals localized near the optimum would suggest an identifiable subset. We use a cutoff of 90\% to determine if parameters are identifiable, i.e., we consider them identifiable if 90\% of the confidence intervals generated include the true, data-generating parameter. 

\subsection{Profile-Likelihood}
Our final approach for assessing identifiability is the gold-standard univariate profile-likelihood \citep{colebank2022silico,colebank_assessing_2025,simpson_parameter_2026,raue_structural_2009}. Similar to the confidence interval approach, the profile-likelihood is used to derive bounds for parameter estimates. However, the profile-likelihood directly assesses these confidence bounds through repeated numerical optimization, rather than depending on the local-sensitivity, first-order approximation of the Fisher-information matrix \citep{simpson_parameter_2026,raue_structural_2009}. 

Given the vector of noisy observations $\bm{Y}=\left\{y_{ij}\right\}$ for time-point $i$ in state $j$ and the corresponding vector of model evaluations $\bm{F}(\bm{\tilde{\theta}})=\left\{f_j(t_i;\tilde{\bm{\theta}})\right\}$, we define the negative log-likelihood as
\begin{equation}\label{eq:loglike}
    -LL(\bm{\tilde{\theta}}) \propto \left(\bm{Y} -\bm{F}(\bm{\tilde{\theta}})) \right)^\top \mathbf{\Sigma^{-1}}\left(\bm{Y} -\bm{F}(\bm{\tilde{\theta}}) \right).
\end{equation}
For this study, the weight matrix, $\mathbf{\Sigma}$, has diagonal entries that represent the mean data for each state $j=1,\dots,s$ from the different experimental designs, $\bar{y}_j$. This makes it equivalent to the WLS problem in eq. \eqref{eq:WLS}.

The univariate profile-likelihood for a parameter $\theta_i$ is then defined as
\begin{equation}\label{eq:PL}
    PL_i(\tilde{\theta}_i) = \min_{\bm{\tilde{\theta}_{\sim i}}} \left[-LL\left(\tilde{\theta}_i,\bm{\tilde{\theta}_{\sim i}} | \bm{Y}\right)\right]
\end{equation}
where $\tilde{\theta}_i$ is the parameter being profiled and $\tilde{\theta}_{\sim i}=\{\tilde{\theta}\} \setminus \tilde{\theta}_i$ is the remaining parameters to be inferred. Univariate confidence intervals can be computed for each parameter by comparing the profile-likelihood to the inverse cumulative distribution function of the chi-square distribution, $\mathcal{X}$
\begin{equation}\label{eq:PL_CI}
    \text{CI}_{\text{PL}}^{1-a}(\theta_i) = \left\{\theta_i | 2\left[LL(\bm{\theta}_{MLE}|\mathbf{y})-PL_i(\theta_i)\right] \leq \chi^2_{1,1-a}\right\}
\end{equation}
where $\bm{\tilde{\theta}_{MLE}}$ is the maximum likelihood estimator (MLE) of the full parameter set corresponding to the maximum likelihood (minimum of the negative log-likelihood). The difference between the profile-likelihood and MLE evaluated log-likelihood are compared to a chi-squared distribution quantile with one-degree of freedom and an $1-a$ confidence level. We use $a=0.05$ in all experiments.

Constructing the profile-likelihood is algorithmically structured as follows: given some initial parameterization, here being a minimizer found through optimization, proceed with the following:
\begin{enumerate}
    \item Select parameter $\tilde{\theta}_i$ as your ``profiled-parameter,'' and then define $\bm{\tilde{\theta}_{\sim i}} = \{\bm{\tilde{\theta}}\} \setminus \tilde{\theta}_i$ as the remaining parameters.
    \item Select a range of values  for $\tilde{\theta}_i$, e.g. $\pm50\%$ of its optimal value. Then solve the optimization problem in eq. \eqref{eq:PL} for each value of $\tilde{\theta}_i$.
    \item Repeat this procedure for all parameters in the subset.
    \item Plot the profile-likelihood and determine $\text{CI}_{\text{PL}}^{1-a}$ at some significance level $1-a$. Parameters that have confidence intervals that \textbf{do not} have values greater than $\mathcal{X}(a)$ are \textbf{non-identifiable}. Parameters that cross this threshold are considered identifiable.
\end{enumerate}

\section{Results}

\subsection{Global Sensitivity}
\begin{figure}[ht!]
    \centering
    \includegraphics[width=0.8\linewidth]{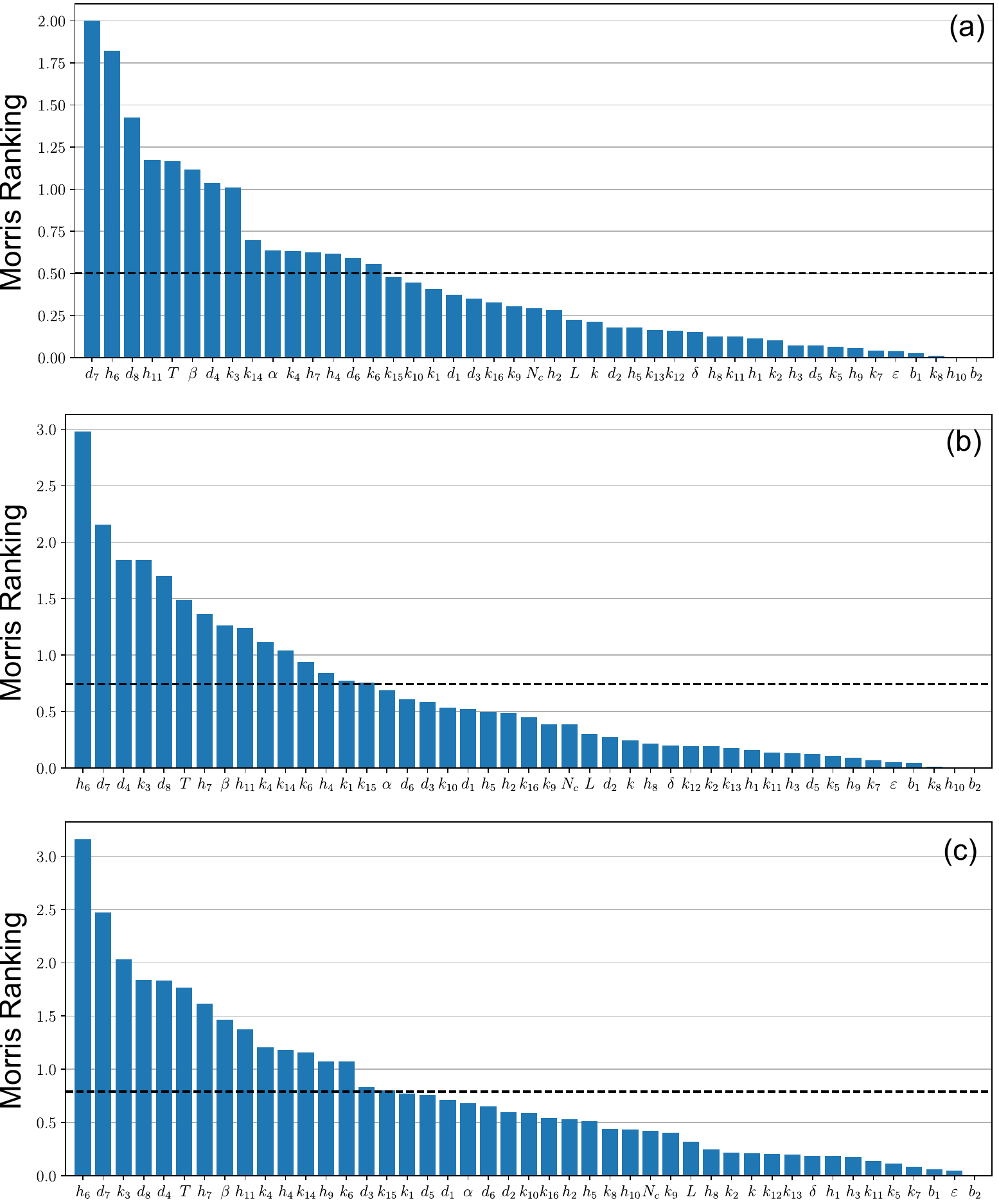}
    \caption{Parameters ranked by bar height of the metric $\mathcal{R}$. Parameters that do not reach the threshold, demarcated by a horizontal dotted line, are considered non-influential a) Ranking for $D_1$ b) Ranking for $D_2$ c) Ranking for $D_3$}
    \label{fig:paramRankings}   
\end{figure}

We computed the Morris Screening sensitivity measures from 100 samples with a step size of $\Delta\approx0.508$. The sensitivities were computed for each model state and then concatenated based on the experimental designs and ranked using the metric $\mathcal{R}$ defined in \cref{eq:rank}. This is visualized in \cref{fig:paramRankings}. In descending order, selecting every parameter with $\mathcal{R}$ above the defined threshold builds the following sets:
\begin{align}
    & \bm{\theta}_{D_1}: \{d_7, h_6, d_8, h_{11}, T, \beta, d_4, k_3, k_{14}, \alpha, k_4, h_7, h_4, d_6, k_6\} \nonumber\\ 
    \label{eq:selectSets}
    & \bm{\theta}_{D_2}: \{h_6, d_7, d_4, k_3, d_8, T, h_7, \beta, h_{11}, k_4, k_{14}, k_6, h_4, k_1, k_{15}\} 
    \\
   & \bm{\theta}_{D_3}: \{h_6, d_7, k_3, d_8, d_4, T, h_7, \beta, h_{11}, k_4, h_4, k_{14}, h_9, k_6, d_3, k_{15}\} \nonumber
\end{align}
These parameters are further analyzed by the three identifiability methods.

\subsection{Local Sensitivity}
Local derivative based sensitivities were computed for the scaled residuals ($e_{ij}$ defined in eq.~\eqref{eq:WLS}) and used to calculate the Fisher information matrix $\bm{F}$. Across the parameter sets $\bm{\theta}_{D^m_1}$, $\bm{\theta}_{D^m_2}$, and $\bm{\theta}_{D^m_3}$, we found that the condition number of $\bm{F}$ did not exceed the set condition number threshold of $10^6$ with all selected parameters in \cref{eq:selectSets} considered. Therefore, the selected sets are not pared down for additional analyses. This effectively implies that our parameters are all ``identifiable'' from this test.

\subsection{Global Optimization}
\begin{sidewaysfigure}
    \centering
    \includegraphics[width= \linewidth]{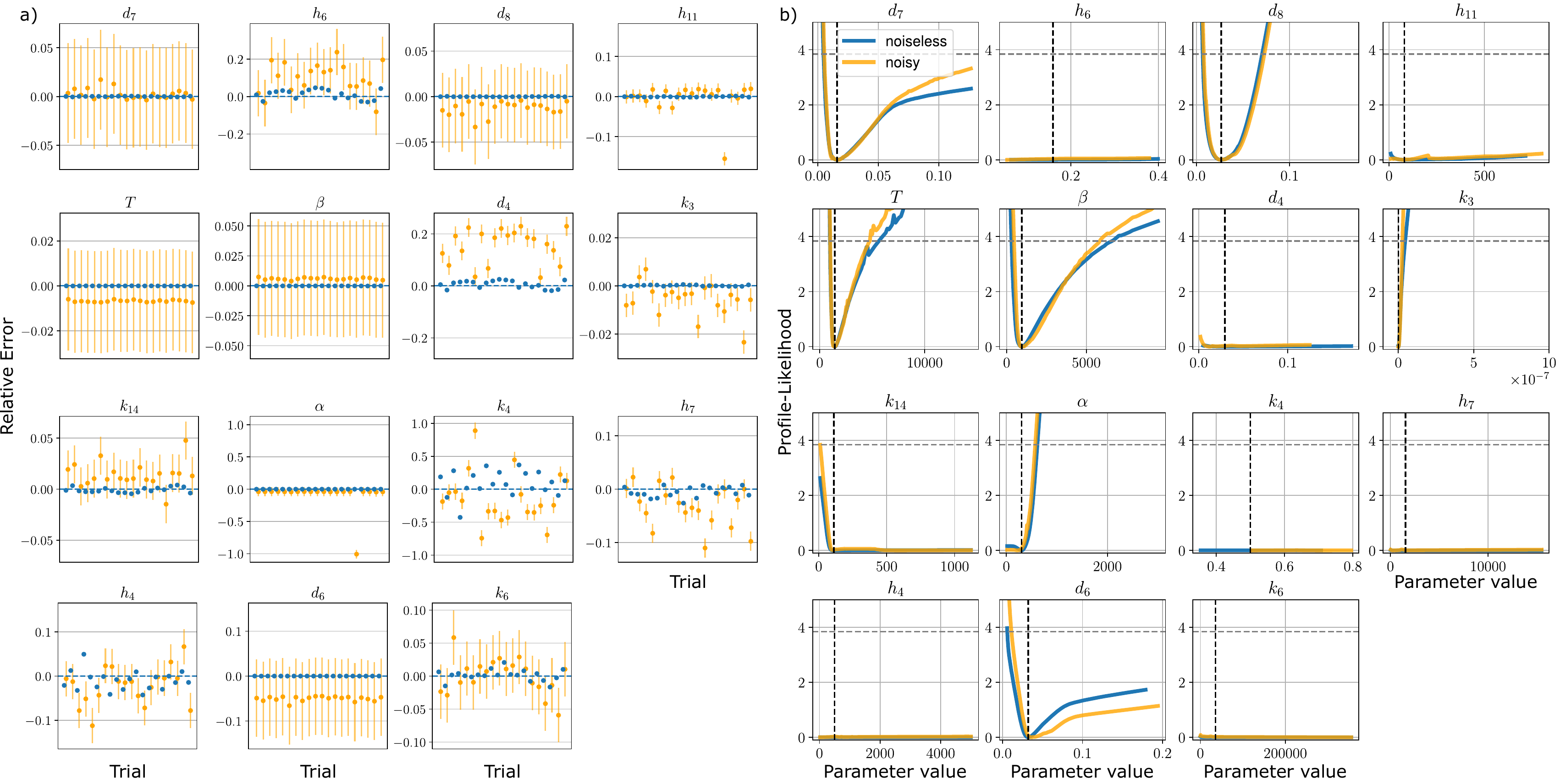}
    \caption{a) Relative error in selected parameter estimates from multi-start inference for $D^{25}_1$. Each of the 20 points across all subgraphs represents a single trial with a new initialization in parameter space. Confidence intervals are represented as vertical lines. The true parameter value is marked by a horizontal dashed line. b) Profile-likelihood curves for each selected parameter, with scaled SSE after optimization plotted against fixed parameter values. Separate curves were calculated for the noise-free and noisy synthetic datasets, and are plotted pair-wise for each parameter. The horizontal dashed lines represent the error bound. The vertical dashed lines represent the respective true parameter value}
    \label{fig:PL_CI_11}
\end{sidewaysfigure}

\begin{sidewaysfigure}
    \centering
    \includegraphics[width= .9\linewidth]{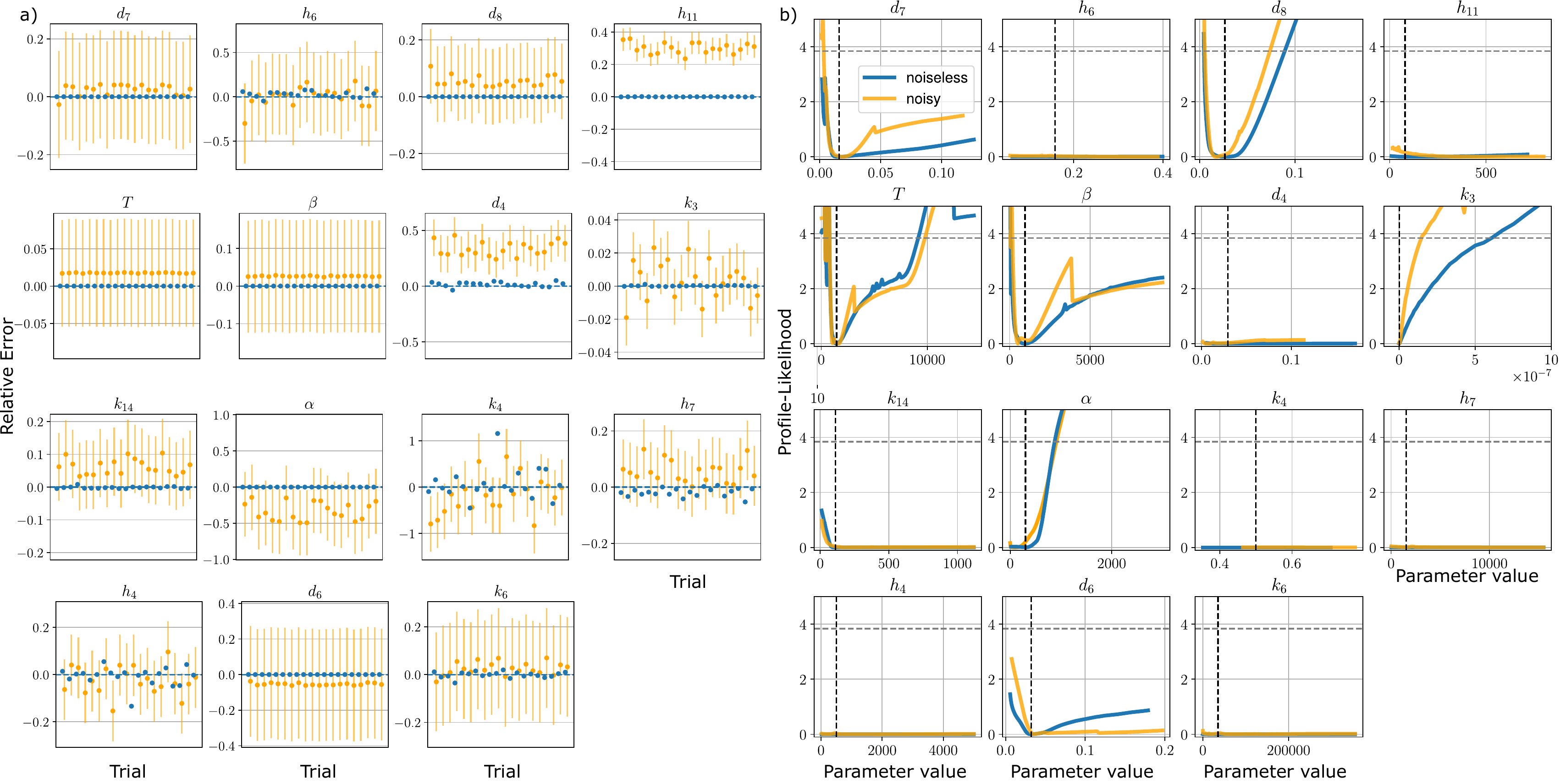}
    \caption{a) Relative error in selected parameter estimates from multi-start inference for $D^{13}_1$. Each of the 20 points across all subgraphs represents a single trial with a new initialization in parameter space. Confidence intervals are represented as vertical lines. The true parameter value is marked by a horizontal dashed line. b) Profile-likelihood curves for each selected parameter, with scaled SSE after optimization plotted against fixed parameter values. Separate curves were calculated for the noise-free and noisy synthetic datasets, and are plotted pair-wise for each parameter. The horizontal dashed lines represent the error bound. The vertical dashed lines represent the respective true parameter value}
    \label{fig:PL_CI_12}
\end{sidewaysfigure}

\begin{sidewaysfigure}
    \centering
    \includegraphics[width= .9\linewidth]{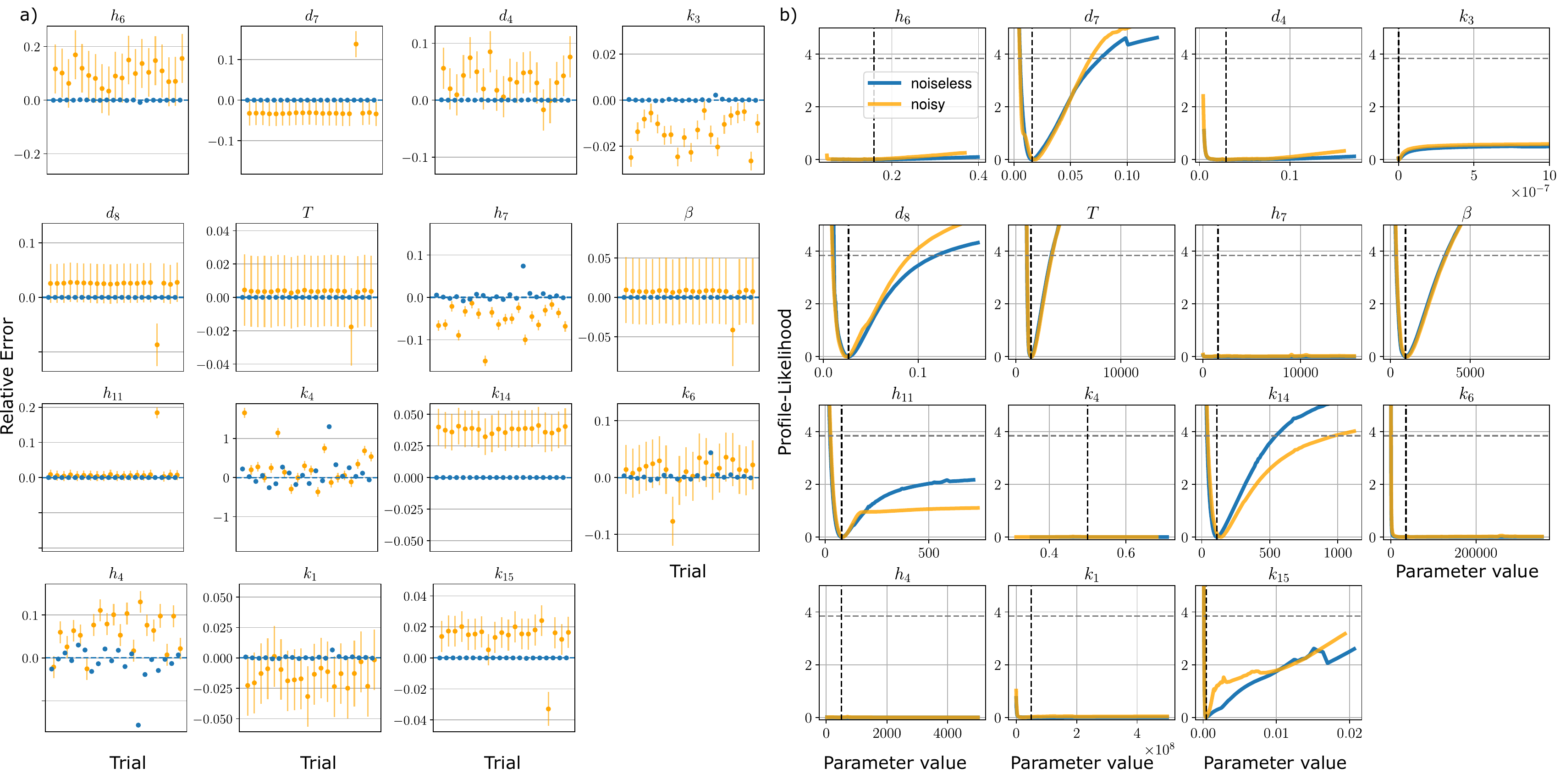}
    \caption{a) Relative error in selected parameter estimates from multi-start inference for $D^{25}_2$. Each of the 20 points across all subgraphs represents a single trial with a new initialization in parameter space. Confidence intervals are represented as vertical lines. The true parameter value is marked by a horizontal dashed line. b) Profile-likelihood curves for each selected parameter, with scaled SSE after optimization plotted against fixed parameter values. Separate curves were calculated for the noise-free and noisy synthetic datasets, and are plotted pair-wise for each parameter. The horizontal dashed lines represent the error bound. The vertical dashed lines represent the respective true parameter value}
    \label{fig:PL_CI_21}
\end{sidewaysfigure}

\begin{sidewaysfigure}
    \centering
    \includegraphics[width= .9\linewidth]{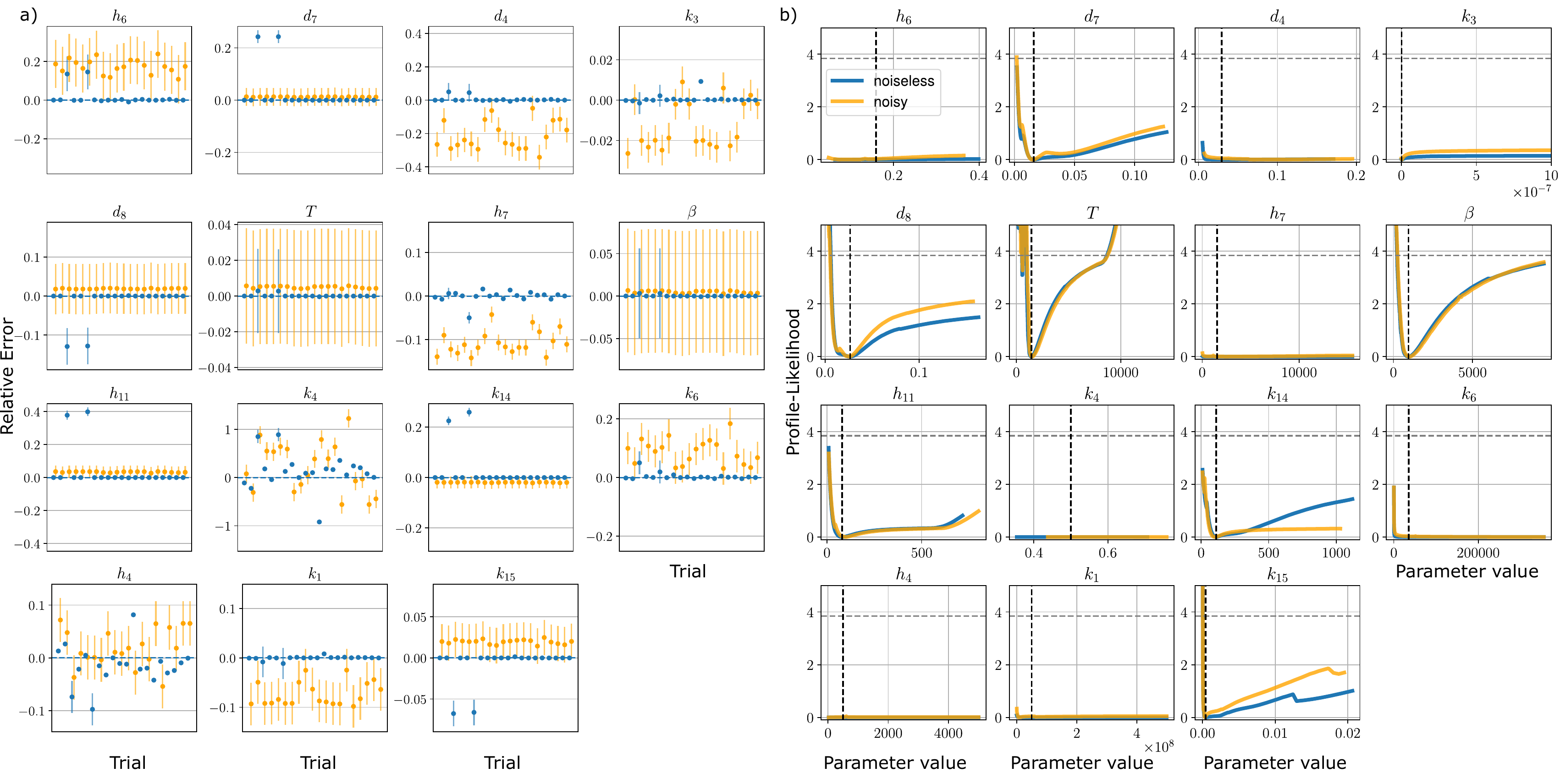}
    \caption{a) Relative error in selected parameter estimates from multi-start inference for $D^{13}_2$. Each of the 20 points across all subgraphs represents a single trial with a new initialization in parameter space. Confidence intervals are represented as vertical lines. The true parameter value is marked by a horizontal dashed line. b) Profile-likelihood curves for each selected parameter, with scaled SSE after optimization plotted against fixed parameter values. Separate curves were calculated for the noise-free and noisy synthetic datasets, and are plotted pair-wise for each parameter. The horizontal dashed lines represent the error bound. The vertical dashed lines represent the respective true parameter value}
    \label{fig:PL_CI_22}
\end{sidewaysfigure}

\begin{sidewaysfigure}
    \centering
    \includegraphics[width= .9\linewidth]{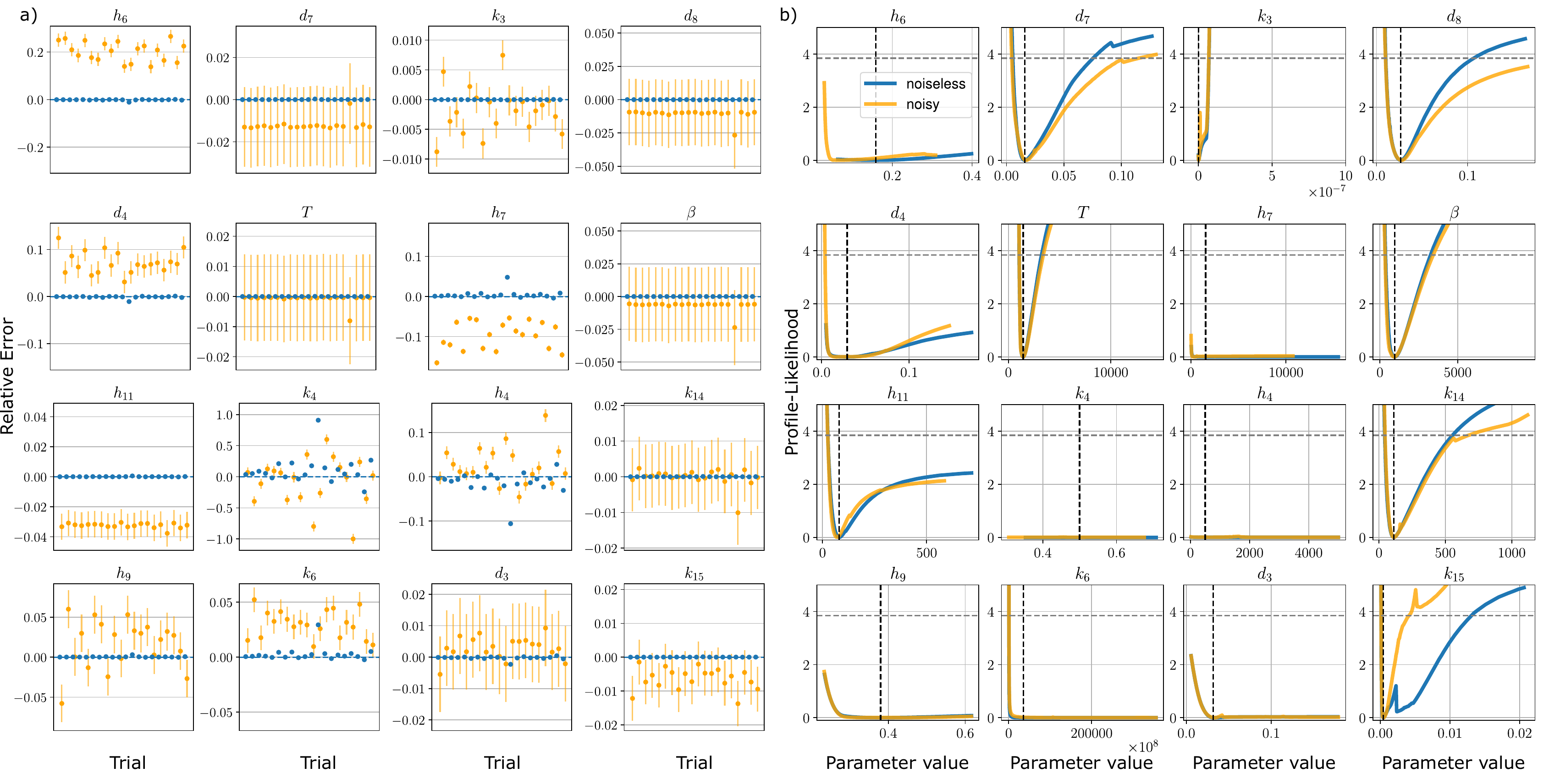}
    \caption{a) Relative error in selected parameter estimates from multi-start inference for $D^{25}_3$. Each of the 20 points across all subgraphs represents a single trial with a new initialization in parameter space. Confidence intervals are represented as vertical lines. The true parameter value is marked by a horizontal dashed line. b) Profile-likelihood curves for each selected parameter, with scaled SSE after optimization plotted against fixed parameter values. Separate curves were calculated for the noise-free and noisy synthetic datasets, and are plotted pair-wise for each parameter. The horizontal dashed lines represent the error bound. The vertical dashed lines represent the respective true parameter value}
    \label{fig:PL_CI_31}
\end{sidewaysfigure}

\begin{sidewaysfigure}
    \centering
    \includegraphics[width= .9\linewidth]{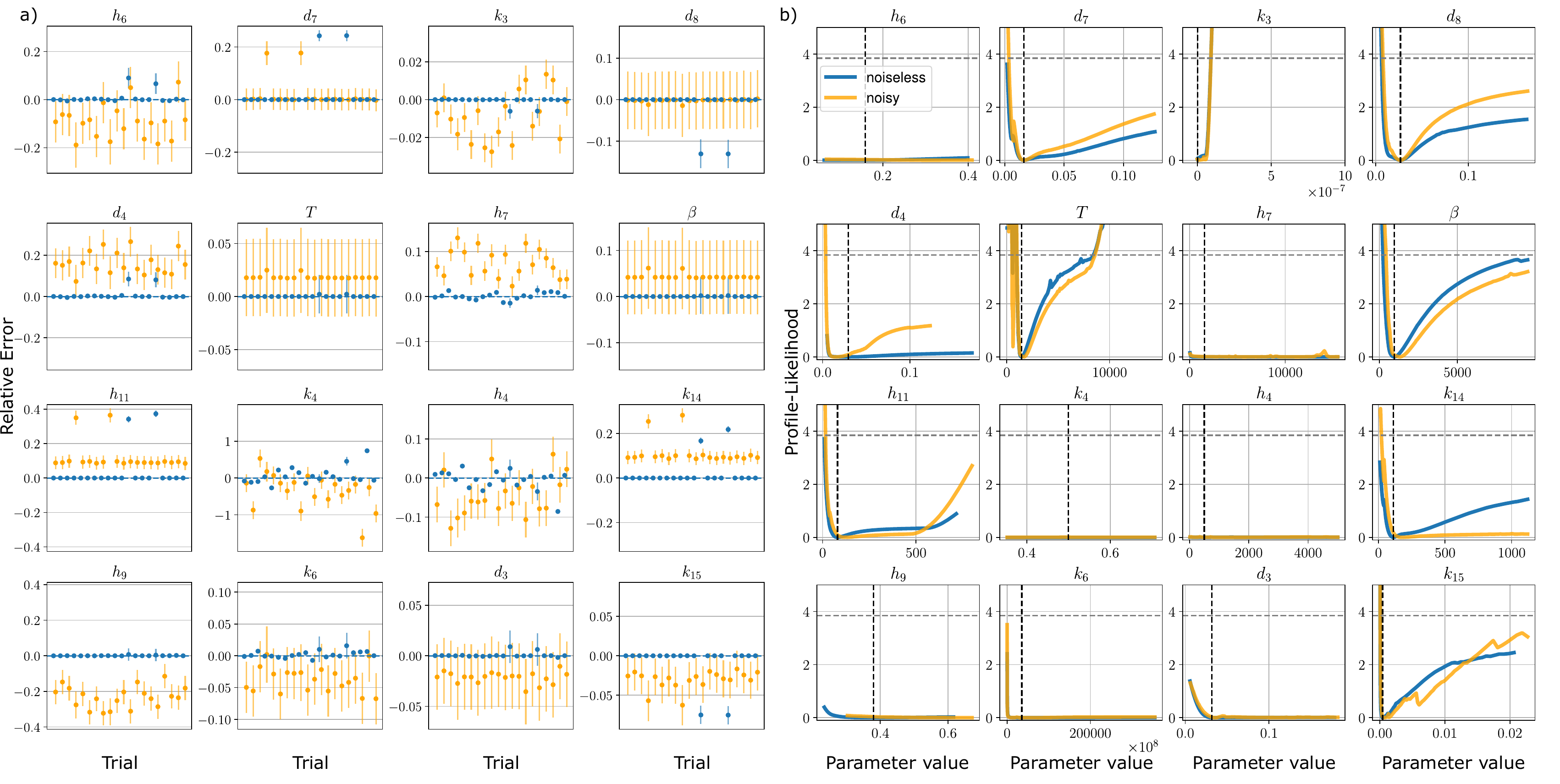}
    \caption{a) Relative error in selected parameter estimates from multi-start inference for $D^{13}_3$. Each of the 20 points across all subgraphs represents a single trial with a new initialization in parameter space. Confidence intervals are represented as vertical lines. The true parameter value is marked by a horizontal dashed line. b) Profile-likelihood curves for each selected parameter, with scaled SSE after optimization plotted against fixed parameter values. Separate curves were calculated for the noise-free and noisy synthetic datasets, and are plotted pair-wise for each parameter. The horizontal dashed lines represent the error bound. The vertical dashed lines represent the respective true parameter value}
    \label{fig:PL_CI_32}
\end{sidewaysfigure}

In our second identifiability assessment, we employed a global optimization routine for both noise-free and noisy data for each experimental design. Each dataset is generated according to the designs $D^m_1$, $D^m_2$, $D^m_3$. Confidence intervals are calculated for all 20 trials. We provide the confidence intervals in \crefrange{fig:PL_CI_11}{fig:PL_CI_32} in the subplot (a) position. The optimal parameter estimate and confidence bounds are presented on a relative scale to make it easier to interpret the variability in the parameters. Values closer to zero represent more accurate parameter estimates. Blue and yellow results represent noise-free and noisy datasets.

We use the confidence interval widths and the locations of the optimal parameter estimates to assess parameter identifiability. For instance, parameter $d_7$ in design $D^{25}_1$ (Figure \ref{fig:PL_CI_11}) appears close to the zero line in the noise-free case, and also has 95\% confidence intervals that contain a relative error of 0 across all 20 initializations spanning $\approx 5\%$. We would interpret this result as suggesting that $d_7$ is identifiable. In contrast, the parameter $d_4$ appears close to the true data generated parameters in $D^{25}_1$ when there is no noise, whereas estimates of $d_4$ and its corresponding confidence intervals lead to insufficient coverage when noise is added in, with errors close to 20\%. There are also borderline results: parameter $k_{14}$ in $D^{25}_1$ consistently appears close to the true parameter value and has confidence intervals that contain the true parameter in 90\% of the initializations. It is unclear whether this parameter is identifiable or not, given the bias away from the true parameter values are around 5\%. 

We now summarize the results for this assessment, and focus our conclusions on the designs that include measurement noise in the data. We can confidently say the following parameters are identifiable based on 90\% of their confidence intervals including the true parameter and their biases falling within a $\pm10\%$ window:
\begin{itemize}
    \item For $D^{25}_1$: $\left\{d_7,d_8,T,\beta,k_{14},\alpha,d_6,k_6\right\}$
    \item For $D^{13}_1$: $\left\{d_7,h_6,d_8,T,\beta,k_3,k_{14},h_7,h_4,d_6,k_6\right\}$

     \item For $D^{25}_2$: $\left\{d_7,d_8,T,\beta,h_{11},k_6,k_1\right\}$
    \item For $D^{13}_2$: $\left\{d_7,d_8,T,\beta,h_{11},k_{14},k_{15}\right\}$

     \item For $D^{25}_3$: $\left\{d_7,d_8,T,\beta,k_{14},d_3\right\}$
    \item For $D^{13}_3$: $\left\{d_7,d_8,T,\beta,d_3\right\}$.
\end{itemize}

\subsection{Profile-Likelihoods}
Profile-likelihood curves were calculated using both noise-free and noisy data for each experimental design. Curves for each parameter are visualized in the right column of \crefrange{fig:PL_CI_11}{fig:PL_CI_32}. We can now compare the identifiability of parameters using this approach versus the confidence interval approach from the previous section. These results suggest the following parameters are identifiable in the case of noisy data for the six designs, with the change in parameters relative to the global optimization approach provided:

\begin{itemize}
    \item For $D^{25}_1$: $\left\{d_7,d_8,T,\beta,\alpha,k_{3}\right\}$, removing parameters $d_6,k_6,$ and $k_{14}$ but adding $k_3$ (we included $\alpha$ since it is identifiable on the side away from the lower bound; we include $d_7$ since it is identifiable with a slightly larger interval). \vspace{0.3cm}
    
    \item For $D^{13}_1$: $\left\{d_8,T,k_{3},\alpha\right\}$, removing parameters $d_7,d_6,k_6,k_{14},\beta,h_4,h_6,$ and $h_{7}$ but adding $\alpha$ (we included $k_3$ and $\alpha$ since they are identifiable on the side away from the lower bound). \vspace{0.3cm}

     \item For $D^{25}_2$: $\left\{d_7,d_8,T,\beta,k_{14}\right\}$, removing parameters $h_{11},k_1,$ and $k_6$ but adding $k_{14}$. \vspace{0.3cm}
     
    \item For $D^{13}_2$: $\left\{T,\beta\right\}$, removing $d_7,d_8,h_{11},k_{14},$ and $k_{15}$ (we included $\beta$ since it is identifiable with a slightly larger interval). \vspace{0.3cm}

     \item For $D^{25}_3$: $\left\{d_7,d_8,T,\beta,k_{14},k_3,k_{15}\right\}$, removing $d_3$ and adding $k_3$ and $k_{15}$ (we included $d_8$ since it is identifiable with a slightly larger interval). \vspace{0.3cm}
     
    \item For $D^{13}_3$: $\left\{T,\beta,k_3,h_{11} \right\}$, removing parameters $d_7,d_8,$ and $d_3$ but adding $k_3$ and $h_{11}$ (we included $\beta$ since it is identifiable with a slightly larger interval). \vspace{0.3cm}
\end{itemize}
These results offer an alternative interpretation of whether the influential parameters can be determined from the experimental designs. We also note that these results represent the noisy datasets. Identifiability in the noise-free setting are mostly consistent with those described above in the noisy data setting, with some minor exceptions like $d_4$ in design $D_3^{25}$ and $D_3^{13}$, where the addition of noise actually makes the parameters identifiable from the left, in contrast to the noise-free case.

\begin{figure}
    \centering
    \includegraphics[width = \linewidth]{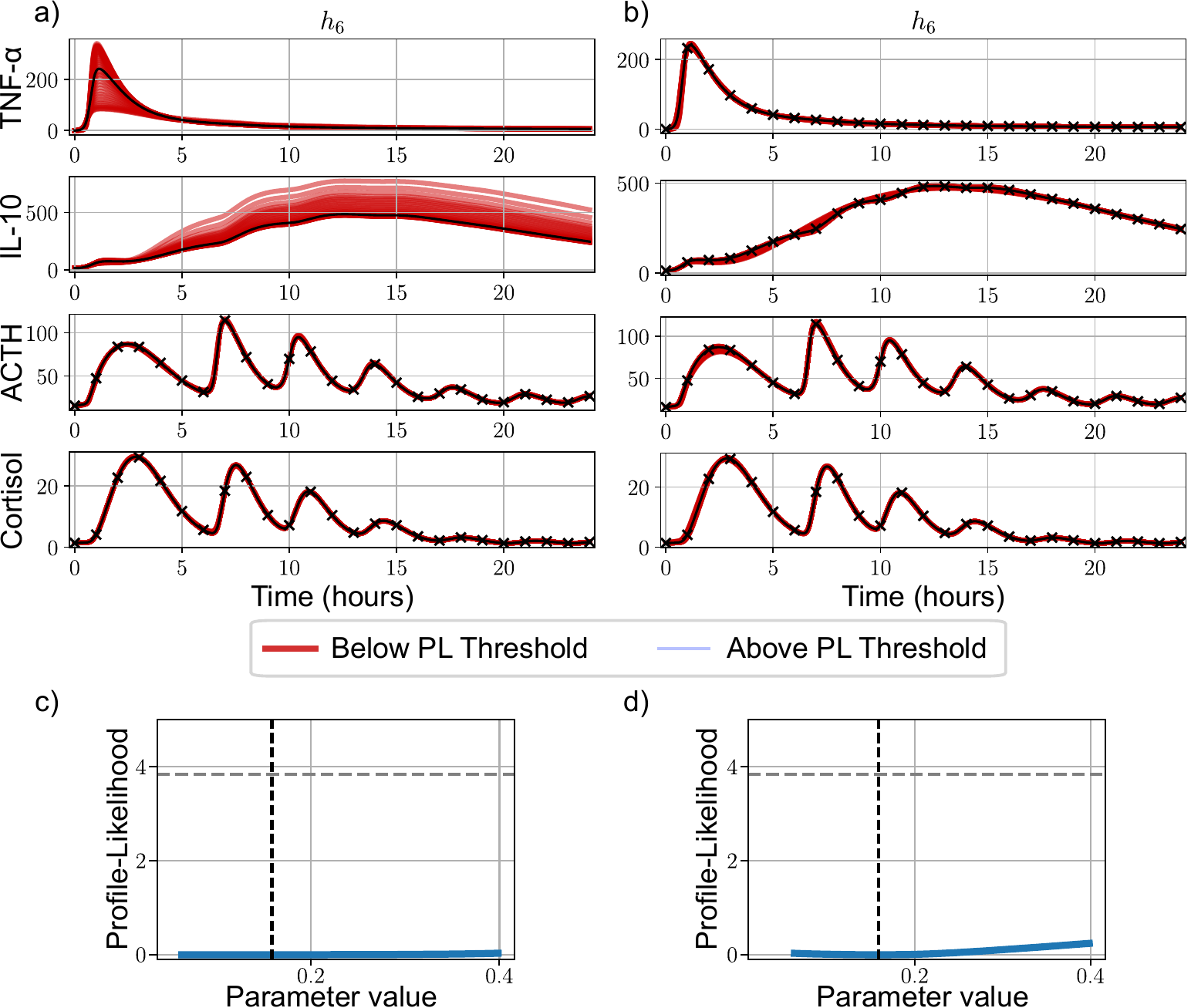}
    \caption{Solution curves where $h_6$ is fixed at a range of different values while the rest of the selected parameters are optimized to noise-free data, as is done for the profile-likelihood procedure. Dark red curves represent solutions where $h_6$ is below the confidence bound, while light blue curves are solutions outside the confidence bounds. The true solution is represented by an opaque black line and data by `x' marks. a) $D^{25}_1$ b) $D^{25}_3$ c) Profile-likelihood for $h_{6}$ in $D^{25}_2$ d) Profile-likelihood for $h_{6}$ in $D^{25}_3$}
    \label{fig:h6_compare}
\end{figure}

\begin{figure}
    \centering
    \includegraphics[width = \linewidth]{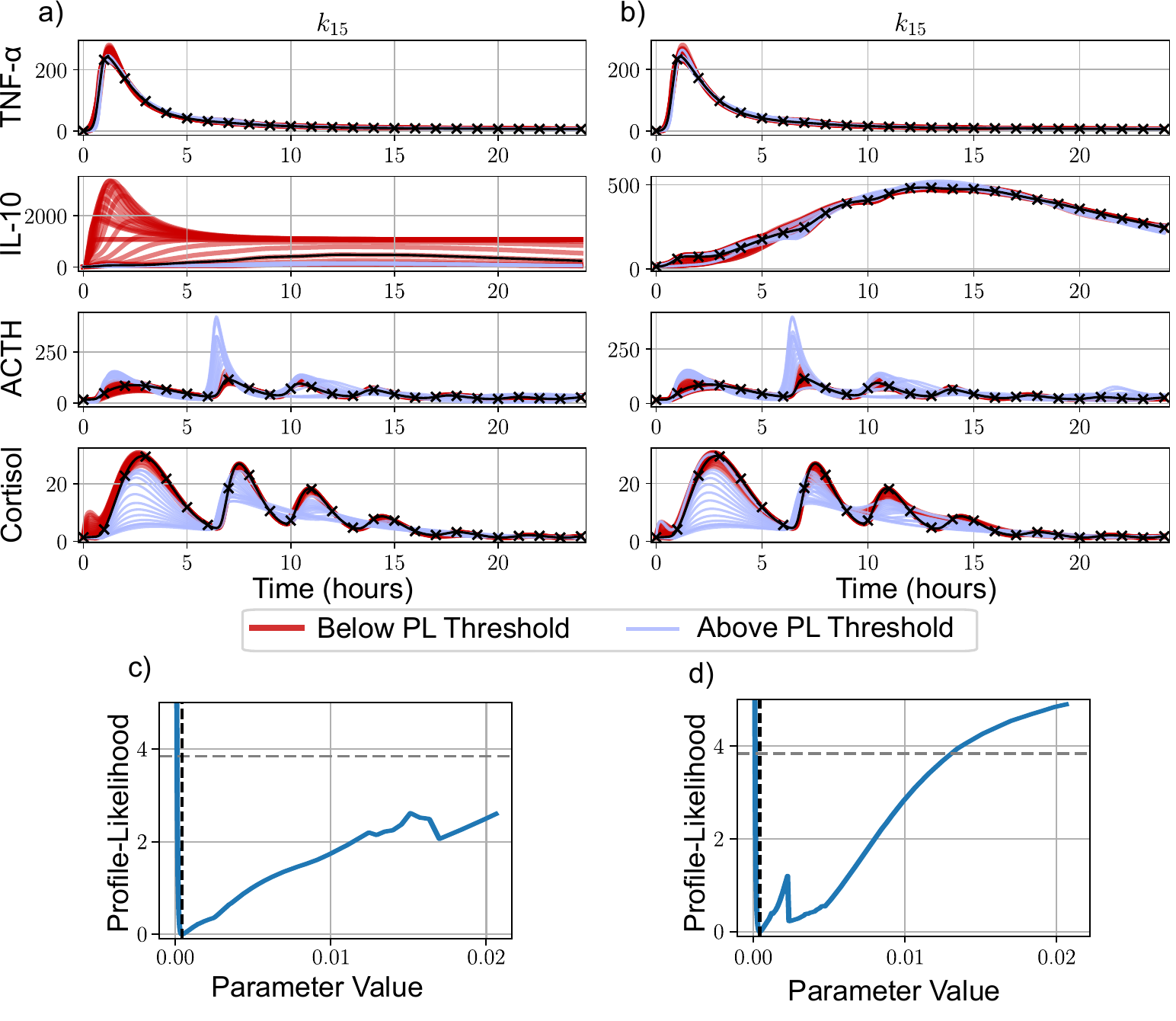}
    
    \caption{Solution curves where $k_{15}$ is fixed at a range of different values while the rest of the selected parameters are optimized to noise-free data, as is done for the profile-likelihood procedure. Dark red curves represent solutions where $k_{15}$ is below the confidence bound, while light blue curves are solutions outside the confidence bounds. The true solution is represented by an opaque black line and data by `x' marks. a) $D^{25}_2$ b) $D^{25}_3$ c) Profile-likelihood for $k_{15}$ in $D^{25}_2$ d) Profile-likelihood for $k_{15}$ in $D^{25}_3$}
    \label{fig:k15_compare}
\end{figure}

\begin{sidewaysfigure}
    \centering
    \includegraphics[width = \linewidth]{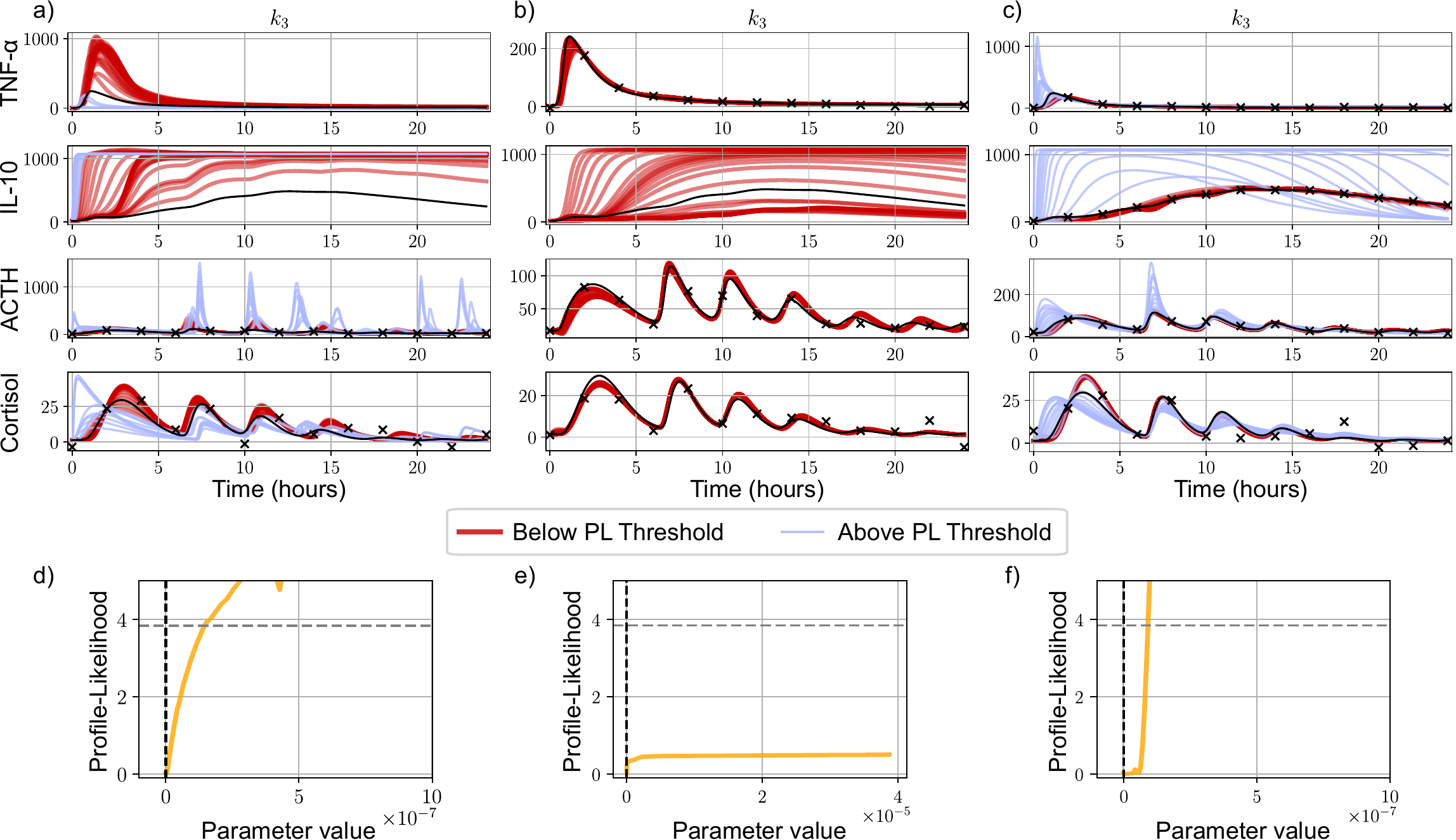}
    
    \caption{Solution curves where $k_3$ is fixed at a range of different values while the rest of the selected parameters are optimized to noisy data, as is done for the profile-likelihood procedure. Dark red curves represent solutions where $k_3$ is below the confidence bound, while light blue curves are solutions outside the confidence bounds. The true solution is represented by an opaque black line and data by 'x' marks. Three designs are investigated with the solution curves on top and the corresponding profile likelihood curve below a) Solution curves for $D^{13}_1$ b) Solution curves for $D^{13}_2$ c) Solution curves for $D^{13}_3$ d) Profile-likelihood for $k_3$ in $D^{13}_1$ e) Profile-likelihood for $k_3$ in $D^{13}_2$ f) Profile-likelihood for $k_3$ in $D^{13}_3$.}
    \label{fig:k3_compare}
\end{sidewaysfigure}

\begin{figure}
    \centering
    \includegraphics[width=0.9\linewidth]{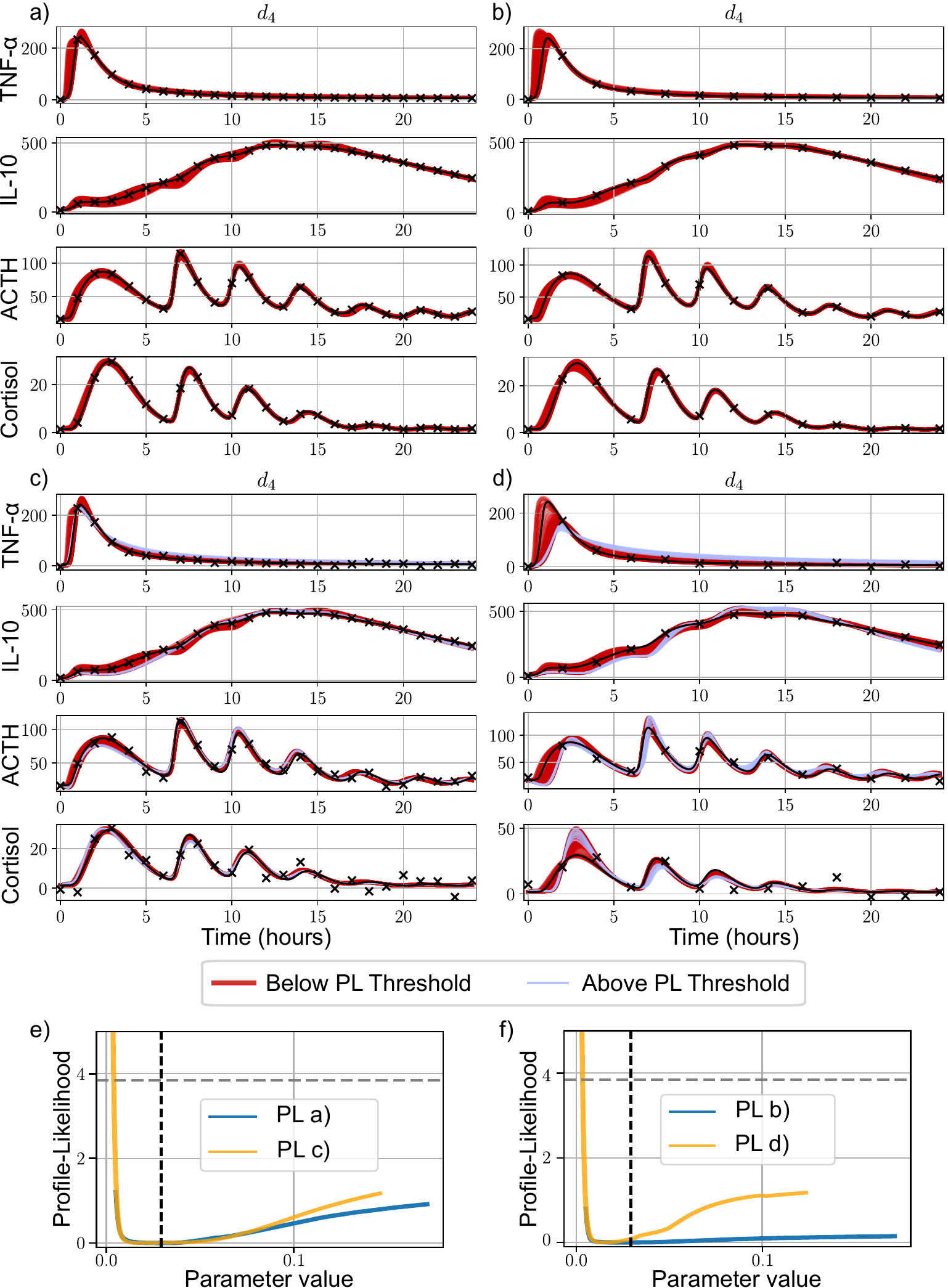}
    \caption{Solution curves where $d_4$ is fixed at a range of different values while the rest of the selected parameters are optimized to noisy data, as is done for the profile-likelihood procedure. Dark red curves represent solutions where $d_4$ is below the confidence bound, while light blue curves are solutions outside the confidence bounds. The true solution is represented by an opaque black line and data by 'x' marks. Results for four designs are plotted: a) $D^{25}_3$ no noise, b)$D^{13}_3$ no noise, c) $D^{25}_3$ noisy, d) $D^{13}_3$ noisy, e) Profile-likelihood for $d_4$ in $D^{25}_3$, and f) Profile likelihood for $d_4$ in $D^{13}_3$}
    \label{fig:d4_compare}
\end{figure}

We can also examine individual parameters more closely by graphing the simulator outputs for each fixed value of the parameter of interest in \crefrange{fig:h6_compare}{fig:d4_compare}. We investigate four different parameters and their behavior. We first present solution curves from values of $h_6$ in \cref{fig:h6_compare} in the noise-free designs $D_1^{25}$ and $D_3^{25}$. The parameter $h_6$ presents with nearly flat profile-likelihood landscapes, and the solutions from the model in the ACTH (denoted by $A$ in the ITIS system) and Cortisol (denoted by $F$ in the ITIS system) states cluster tightly around the observations in these two states. This by definition leads to non-identifiability of this parameter, even though solutions vary considerably around unobserved states in \cref{fig:h6_compare}(a). This is expected, as none of the parameters are being informed by these states and therefore we do not apply them in our determination of identifiability. However, as we calculate the profile-likelihood for $h_6$ in design $D_3^{25}$, we see that the parameter provides nearly identical solutions across all four observed states. This again illustrates why this parameter is non-identifiable: given this range of parameter values for $h_6$, the remaining parameters can compensate to provide equally good predictions to the data.

We then investigate $k_{15}$, which is a parameter that sits on the threshold of practical identifiability between designs $D_2^{25}$ and $D_3^{25}$. Here we focus on the noise-free setting. When only 3 states are observed in $D_2$, the parameter does not cross our statistical threshold for increasing values of $k_{15}$. The predictions in \cref{fig:k15_compare}(a) show how small values of $k_{15}$ lead to solutions far from the data, whereas larger values lead to a collection of predictions in ACTH, Cortisol, and TNF-$\alpha$ (states $A,F,$ and $T$) that are indistinguishable. Predictions of IL-10 (state $I$), which is not observed in $D^{25}_2$, are drastically different along this profile-likelihood. Adding observations for IL-10 in $D^{25}_3$ restores identifiability for the parameter. The solution curves for the original 3 observed states do not change significantly, while the fit to IL-10 becomes much stronger. The solutions around TNF-$\alpha$ and IL-10 (when observed) appear clustered tightly compared to ACTH and cortisol.

The parameter $k_3$ presents an interesting property: it is identifiable by the profile-likelihoods for designs $D^{13}_1$ and $D^{13}_3$, but is  non-identifiable in $D^{13}_2$. The individual solutions in \cref{fig:k3_compare} corroborate this conclusion. Optimized solutions vary considerably for $D_1$ and $D_3$ when $k_3$ is identifiable, but cluster together in $D_2$. Note that the range of values tested in $D_2$ is also considerably larger than the other two, suggesting even more uncertainty in parameter estimates with this design. The option to optimize to data for TNF-$\alpha$ impacts the parameter space in a way that allows the optimal solution to be found by changing other parameters, regardless of $k_3$'s value. Introducing observations for IL-10 then reverses this effect.

Finally, we present results for a parameter that shows slightly improved identifiability when noise is present and when the number of samples are reduced. Results for $d_4$ in \cref{fig:d4_compare} suggest that the parameter is practically non-identifiable in any model design. However, we find that within $D_3$, the parameter performs best when data is reduced and noise is added. Neither noise-free datasets lead to finite confidence bounds, while the noisy datasets provide bounds on the left. 


\section{Discussion}
This work provides a detailed comparison of identifiability methods for a system of nonlinear ODEs, which have been used by several authors to investigate the interactions between the neuorendocrine and inflammatory system \citep{bangsgaard_integrated_2017,chow_acute_2005,dobreva2021physiological}. The goal of this work is to provide a critical comparison of identifiability workflows for different observation operators. Identifiability assessment is critical in the biological and medical modeling space, as inferred parameters can be correlated with or compared to actual biomarkers of underlying processes. We compared a common, computationally affordable option (global sensitivity analysis followed by a local analysis using the Fisher information matrix) to two more complex assessments that leverage either (a) multistart optimization and asymptotic-derived confidence intervals or (b) the profile-likelihood.

\subsection{Observation Operator and Global Sensitivity}
From a strictly modeling and inference perspective, we analyze multiple experimental designs for which data may be available. This is driven in part by the fact that different users of a computational model (e.g., hospital systems or researchers looking into inflammation) may have access to different measurements and hence observation operators may change. For example, the modeling study presented in \citep{dobreva2021physiological} uses experimental data from several inflammatory markers at different time intervals during an endotoxin challenge \citep{Janum2016,Copeland2005}. In contrast, the original ITIS model developed in \citep{bangsgaard_integrated_2017} used a separate dataset during endotoxin challenge that recorded a different set of inflammatory states, but at regular 30 minute intervals \citep{Clodi2008}. This shows that a single model, such as the ITIS model constructed by Bangsgaard et al., might be repurposed in multiple ways depending on the available observations, timing of observations, and noise levels. This serves as a major motivation for our approach.

Global sensitivity analysis is a gold standard for assessing which parameters have the largest impact on a system \citep{smith_uncertainty_2014}. These methods include those leveraging statistical moments (e.g., variance-based sensitivity analysis \citep{colebank_assessing_2025}) or those looking at more empirical or moment-independent measures \citep{smith_uncertainty_2014}. Morris screening is more affordable than other global sensitivity methods in part because it is assessing parameter importance at a coarse scale. The method has been successful in many biological applications where the parameter dimensionality is moderate to large \citep{colebank2022silico,olsen_parameter_2019,Wentworth2016}. We choose our quantity of interest to be the full 24-hour trajectory of each individual state, and then use elementary effects to help identify influential parameters when the experimental design changes. Typically sensitivity analyses are either done for a single state variable or for a specific objective function, such as what we employ in eq. \eqref{eq:WLS} \citep{colunga_parameter_2023,olsen_parameter_2019,dobreva2021physiological}. We abstained from doing this explicitly since the specifics of the time-point measurements may fluctuate, reflecting a more ``general'' sensitivity analysis that focuses on observable states, rather the exact experimental design. Our results contrast those of the original ITIS manuscript, which used local sensitivities to determine the most influential parameters. In contrast to these prior studies, we explicitly include the circadian rhythm parameters in our analysis, since these in principal could be different among subjects. We see later in our identifiability analyses that the ranking order of which parameters are most influential does not appear to be an indicator of identifiability. This is expected since sensitivity analyses are best suited for finding non-influential parameters only. We find some parameters with the lowest rankings, such as $k_6$ and $d_6$ in design $D_1$, $k_1$ and $k_{15}$ in design $D_2$, and $d_3$ in design $D_3$, are identifiable. In contrast, the most influential parameters, like $h_{11}$ in $D_1$ and $h_6$ in $D_2$ and $D_3$, are not identifiable.

\subsection{Sensitivity Based Identifiability}
Our first methodology for assessing parameter identifiability is local sensitivity analysis and the construction of the Fisher information matrix. This is often the only identifiability analysis used, given its lower computational complexity and relatively close tie to frequentist statistics \citep{smith_uncertainty_2014}. While analyzing the Fisher information alone can help establish identifiability \citep{dadashova_local_2024,cintron-arias_sensitivity_2009,wang_systematic_2025}, several authors have also transformed the Fisher information into an approximate covariance and then correlation matrix \citep{colunga_parameter_2023,dobreva2021physiological,olsen_parameter_2019}. Both these methods provide information about identifiability in the small neighborhood around the parameter. However, these methods can be difficult to interpret when considering what magnitude of the condition number is deemed ``computationally singular.'' We opted to use a threshold of $10^6$ in this work, though other studies have used other thresholds \citep{colebank2022silico}. Our analyses using these methods suggested that all parameters in the subsets were identifiable, which conflicts with the results of the global optimization and profile-likelihood analyses that suggest a substantially smaller subset. A quick retrospective investigation of the Fisher-information matrix analysis reveals that lowering the threshold for the condition number (e.g., to $10^4$) would not substantially change the number of parameters deemed ``identifiable.'' 

The global optimization approach is a brute force way of assessing the landscape of the objective function to determine if there are unique cost function values corresponding to unique parameter sets. This was used in the prior study by \citep{colunga_parameter_2023} in the context of cardiovascular modeling, where the variability in the parameter estimates was used to deduce whether parameters were identifiable. The review by Villaverde et al. provides an in depth description of how optimization methods can be used for such analyses (\citeyear{villaverde2019benchmarking}). We innovate on this approach by explicitly calculating parameter confidence intervals using the Fisher information matrix and derived covariance \citep{smith_uncertainty_2014,cintron-arias_sensitivity_2009}. Expanding the Fisher information matrix to a covariance approximation enables the use of the Cram\'er-Rao lower bound for the parameter confidence intervals \citep{miao2011identifiability,simpson_parameter_2026}, which in practice is easier to interpret than the condition number of the Fisher information matrix alone. The results in \crefrange{fig:PL_CI_11}{fig:PL_CI_32} are still difficult to fully interpret, but provide a better visual understanding of uncertainty in the parameter estimates. Our findings show a somewhat paradoxical finding: a more complex observation operator and more observed states tends to reduce the number of identifiable parameters. This has been speculated in prior studies \citep{miao2011identifiability,smith_uncertainty_2014,eisenberg2014determining}, yet this has recieved little attention in mathematical biology. This is especially important when trying to translate models between experimental designs, as results for identifiability appear here to be very problem specific. We do find some consistency across designs, namely that each parameter subset includes the parameters $d_7,d_8,T$, and $\beta$. We conclude that, while constructing multiple confidence intervals doesn't make identifiability abundantly clear, it does provide a significant advantage in trying to interpret uncertainty attributed to model parameters. We can consider it as a screening tool: if a parameter is almost always estimated far from it's true value, with confidence intervals that do not include the true value, and the quality of fit to data is comparable, then there is evidence that a unique parameter value cannot be determined for a given design. However, as was the case for determining the numerical threshold for the Fisher information matrix, determining `almost always' in the global optimization screening needs to be defined for each problem.

\subsection{Profile-likelihood Analyses}
Our final method for analysis was the profile-likelihood, which explicitly computes the parameter confidence intervals point-by-point using optimization. This methodology is largely considered a gold-standard due to its ability to handle nonlinearities and asymmetric shapes for parameter confidence intervals \citep{simpson_parameter_2026}. Several authors have shown distinct differences between the profile-likelihood and other identifiability methods, such as the Fisher information matrix \citep{colebank2022silico,Preston2025,wieland_structural_2021}, which echo our results: the Fisher-information matrix can be overly confident in which parameters are identifiable in comparison to the profile-likelihood. The experimental design that provides the largest set of identifiable parameters is $D_3^{25}$, which only includes seven parameters. The rest of the designs result in smaller parameter subsets. We consistently see that there are more parameters when there are 25 time points versus 13 time points. This is in contrast to what we found with the global optimization and confidence interval approach, where designs $D_1^{13}$ and $D_2^{13}$ provided more identifiable parameters than their 25 time point equivalents. Consistent with both the global optimization and profile-likelihood approaches are the observation that more observable states don't necessarily increase the identifiability of the parameters. This suggests that the inherent interactions between model states can have a critical impact on the ability to infer model parameters. For instance, the parameter $k_3$ is identifiable when observations of ACTH and Cortisol (states $A$ and $F$) are available in $D_1$, but no longer identifiable when we add in observations of TNF-$\alpha$ (state $N$) in $D_2$. This is then remedied with additional observations of IL-$10$ (state $I$). This suggests that there is strong interaction between states that can lead to identifiability issues, especially while estimating other parameters in the system.

Given that these results are difficult to understand in parameter space, we explored the landscape of the simulated outputs in \crefrange{fig:h6_compare}{fig:d4_compare}. We find that $h_6$ (which is in eq.~\eqref{eq:ITIS_N}) is non-identifiable in designs $D_1^{25}$ and $D_3^{25}$. Propagating solutions through the ODE system as shown in \cref{fig:h6_compare} confirms this: when only ACTH and Cortisol data are available, we get nearly identical solutions for these two states, while the solutions for the non-observed states IL-$10$ and TNF-$\alpha$ vary. This latter point may suggest that having measurements in these two states would correct this issue; however, we see that observations in $D_3^{25}$ do not circumvent the identifiability issue. This suggests that there are underlying parameter interactions that can compensate for various values of $h_6$, even though it is highly influential, especially for $D_2$ and $D_3$. This again reinforces that influential parameters can be non-identifiable for a given observation operator and experimental design. Our example using the parameter $k_{15}$ provides a straight-forward case: $k_{15}$ is nearly identifiable when ACTH, Cortisol, and TNF-$\alpha$ are observed, and predictions vary widely for the former two states (but not TNF-$\alpha$). Here we see that our identifiability criteria depend on all three states, and even though ACTH and Cortisol dynamics are clearly unique, there are multiple solutions for TNF-$\alpha$ which overlap. The addition of IL-$10$ in $D_3$ pushes our criteria above the threshold. 

Our last two investigations provide some of the most interesting findings. In \cref{fig:k3_compare}, we generate solutions as we profile $k_3$. This parameter is more influential in designs $D_2$ and $D_3$ than $D_1$ (see \cref{fig:paramRankings}). We observe the following: in design $D_1$ (\cref{fig:k3_compare}(a)) predictions along the profile-likelihood lead to identifiable solutions across all four model states, even though only ACTH and Cortisol are used for inference. When we analyze the model using data in $D_2$, we see a drastic shift: now, $k_3$ is not identifiable, and predictions for TNF-$\alpha$ are tight around the measured data, along with ACTH and Cortisol, along the profile-likelihood. In both $D_1$ and $D_2$, we observe that IL-$10$ has variable predictions. Finally, once we introduce observations for IL-$10$, we regain identifiability and see that predictions outside the confidence interval are drastically different and fall far from the observations. This behavior is not intuitive, but is attributed to the interactions between state variables. Looking at our system of eqs.~\eqref{eq:ITIS_E}-\eqref{eq:ITIS_F}: $k_3$ plays a role in the dynamics of TGF-$\beta$ ($G$), which interacts nonlinearly with IL-$10$ ($I$), TNF-$\alpha$ ($T$), and Cortisol ($F$). The state TNF-$\alpha$ is directly linked to CRH ($C$) (which causes changes in ACTH) and ACTH ($A$). When we only observe ACTH and Cortisol, there is a direct link to Cortisol dynamics, and thus makes identifying this parameter easier. When TNF-$\alpha$ is added, we have nonlinear interactions that are now observed, which may contribute to the non-identifiability we see in $D_2$. We also note that the parameter $k_1$ is influential in $D_2$, yet this parameter scales a majority of the phagocyte ($P$) equation, and $k_3$ interacts with TGF-$\beta$ ($G$) through the term $k_3P$. This likely contributes to the identifiability issue in $D_2$. Finally, since the state $G$ is linked directly to IL-$10$ as well, we get that observations for this additional state provide a constraint that then places our parameter as identifiable.

We conclude with some observations about noise-free versus noisy-data and the impact of sub-sampling of time series data, as illustrated in \cref{fig:d4_compare}. These results all use design $D_3$ and observe four state variables, and we note that parameter $d_4$ is the fifth most influential parameter on this design. If we focus on the case when there is no noise added to the signals (\cref{fig:d4_compare}(a) and \cref{fig:d4_compare}(b)) but we move from more temporal information to less temporal information, we get a slightly flatter profile-likelihood. This indicates, as expected, that parameter confidence shrinks when less data are available. In contrast, when we introduce noise in our observations (\cref{fig:d4_compare}(c) and \cref{fig:d4_compare}(d)), we find that the profile-likelihood values increases, suggesting that there are better constraints on the optimal solution when noise is added. Moreover, it appears that less frequent observations in $D_3^{13}$ provide stronger confidence bounds than $D_3^{25}$ when noise is added. Considering this, we conclude that noise itself is useful in increasing the practical identifiability of a parameter in this case, since it necessarily decreases the possible likelihood values relative to a noise-free dataset. This is not a generalizable statement, though. As observation errors increase, we expect to see flatter likelihoods, as there are minimal differences between the ``optimal'' solution and those found along the profile-likelihood. Nevertheless, this study provides some evidence that a small or moderate amount of noise, as expected in biological data, can improve the numerical values for practical identifiability calculation.

\subsection{Limitations}
There are multiple limitations or points of improvement for our present study. First, we considered a sensitivity-driven parameter subset selection for all of our identifiability analyses. We did this to mimick prior studies in mathematical biology, yet several authors have considered alternative approaches, such as conducting identifiability or uncertainty analyses first \citep{cogan2026order} or selecting physiologically important parameters prior to conducting sensitivity analyses \citep{bangsgaard_integrated_2017}. We only focus here on practical identifiability, given the emphasis on measurement operators, while structural identifiability is not considered. We considered noise-free and Gaussian independent noise models for our experimental designs here. This inherently limits our analysis in part because noise may be multiplicative \citep{simpson_parameter_2026} or there may be model discrepancy present \citep{kimpton2025challenges}. This latter point has received little attention, and requires further mathematical analyses. Finally, we opted for relatively simple experimental designs for our observations and observation operators. A more exhuastive set of designs could be examined in depth, and the use of specific information criteria, the expected information gain for example \citep{liepe2013maximizing}, could directly identify the best observation operator. This would require a significant increase in computational costs to screen various designs and within these identify parameters most influential and identifiable. A more detailed subset selection procedure, such as repeated profile-likelihood with iterative subset reduction, would be robust in this setting. This would require substantially more computation time, though. As noted in \citep{Preston2025} and implemented in \citep{colebank_assessing_2025}, surrogate models and emulation can provide a solution to this bottleneck, and should be considered moving forward.

\backmatter

\section*{Declarations}

\bmhead{Acknowledgements}
The authors gratefully acknowledge the computational resources provided by the Theia high performance computing cluster at the University of South Carolina which is supported by National Science Foundation Major Research Instrumentation Grant No. 2320292. We also acknowledge the technical assistance and resources provided by Research Computing at the University of South Carolina (RRID:SCR\_027488).

 \bmhead{Funding}
 This work was funded in part by a South Carolina EPSCoR Grant for Applications in Industry and Networking Program (26-GA01), through the University of South Carolina’s NIAID-funded R25AI164581 (R25) BigDataHealth Science Fellow Program, and through an American Heart Association Career Development Award (26CDA1599801, doi:10.58275/AHA.26CDA1599801.pc.gr.243168).
 \bmhead{Conflicts of Interests} The authors declare that there are no conflicts of interest.
 \bmhead{Availability of Data and Materials}
 The code and synthetic data used to generate all results is hosted online and openly available in Github \url{https://github.com/aayres22/ITIS_Analysis.git} and Zenodo  \url{https://doi.org/10.5281/zenodo.22773537}\citep{ayres_2026_22773537}. 
 

\bibliography{refs}







\end{document}